\documentclass[a4paper,twocolumn,superscriptaddress,11pt,accepted=2019-06-10]{quantumarticle}
\pdfoutput=1
\usepackage[utf8]{inputenc}
\usepackage[english]{babel}
\usepackage[T1]{fontenc}
\usepackage{amsmath}
\usepackage{hyperref}

\usepackage{tikz}
\usepackage{lipsum}
\usepackage[numbers,sort&compress]{natbib}
\usepackage[utf8]{inputenc}
\usepackage[english]{babel}
\usepackage[T1]{fontenc}
\usepackage{amsmath}
\usepackage{hyperref}

\usepackage{amsmath,amsthm,amssymb,dsfont}
\usepackage{mathtools}
\usepackage{braket}
\newcommand{\ketbra}[2]{| #1 \rangle \langle #2 |}
\newcommand{\evec}[2]{\vec{\bf{W}}_{#1}^{\ [#2]}}
\newcommand{\pvec}[2]{\vec{\bf{p}}_{#1}^{\ [#2]}}
\newcommand{\norm}[1]{\lVert #1 \rVert_1}
\newcommand{\ave}[1]{\langle  W_{#1} \rangle}
\newcommand{\ord}[1]{\mathcal{O}\left( #1 \right)}
\newcommand{\avw}{\langle  W \rangle}
\newcommand{\e}{\mathrm{e}}
\DeclareMathOperator{\Tr}{Tr}

\begin{document}
\title{Imperfect Thermalizations Allow for Optimal Thermodynamic Processes}
\date{\today}

\author{Elisa B\"aumer}
\email{ebaeumer@itp.phys.ethz.ch}
\orcid{0000-0003-1882-8693}
\affiliation{Institute for Theoretical Physics, ETH Zurich, Wolfgang-Pauli-Str.\ 27, 8093 Z\"urich, Switzerland}

\author{Mart{\'i} Perarnau-Llobet}
\orcid{0000-0002-4658-0632}
\affiliation{Max-Planck-Institut f\"ur Quantenoptik, Hans-Kopfermann-Str.\ 1, D-85748 Garching, Germany}

\author{Philipp Kammerlander}
\orcid{0000-0002-1379-3092}
\affiliation{Institute for Theoretical Physics, ETH Zurich, Wolfgang-Pauli-Str.\ 27, 8093 Z\"urich, Switzerland}

\author{Henrik Wilming}
\orcid{0000-0002-0306-7679}
\affiliation{Institute for Theoretical Physics, ETH Zurich, Wolfgang-Pauli-Str.\ 27, 8093 Z\"urich, Switzerland}

\author{Renato Renner}
\orcid{0000-0001-5044-6113}
\affiliation{Institute for Theoretical Physics, ETH Zurich, Wolfgang-Pauli-Str.\ 27, 8093 Z\"urich, Switzerland}

\maketitle

\begin{abstract}
	Optimal (reversible) processes in thermodynamics can be modelled as step-by-step processes, where the system is successively thermalized with respect to different Hamiltonians by an external thermal bath.  However, in practice interactions between system and thermal bath will take finite time, and precise control  of their interaction is usually out of reach. 
	Motivated by this observation, we consider finite-time  and uncontrolled operations between system and bath, which result in thermalizations that are only partial in each step. 
We show that  optimal  processes can still be achieved for any non-trivial partial thermalizations at the price of increasing the number of operations, and characterise the corresponding tradeoff. We focus on work extraction protocols and show our results in two different frameworks: A collision model and a model where the Hamiltonian of the working system is controlled over time and the system can be brought into contact with a heat bath.
Our results show that optimal processes are robust to noise and imperfections in small quantum systems, and can be achieved by a large set of  interactions between system and bath.  
\end{abstract}

\section{Introduction}

Recent years have experienced a renewed interest in understanding thermodynamics in 
the quantum regime~\cite{Goold2016,Vinjanampathy2016}. 
A common assumption in the field is that a system thermalizes to a Gibbs  state when put in contact  with a thermal bath. 
While this is perfectly reasonable, it requires some implicit assumptions: For example,  that a sufficiently long time is available to ensure full thermalization of the system  \cite{Breuer2002}; or, within  approaches based on repeated interactions between system-bath,  that  control is available on the system-bath interaction \cite{ziman2005description}.
The goal of this article is to study how optimal thermodynamic processes are modified when one does not have access to such ``perfect thermalizations".
Our results thereby contribute to a recent research line which aims at understanding the level of control required in thermodynamic protocols \cite{ResourceTheory,Wilming2016,Lekscha2016,Brown2016,Friis2017,Lostaglio2016,Perry2016}.
To ensure that our findings are not particular to a specific model for thermodynamic transformations, we consider two different frameworks in this article. 

Before giving an overview of the  specific frameworks considered, let us briefly explain the main insight of this article. We consider discrete processes consisting of $N$ small steps following a smooth curve between two fixed points in parameter space, where the system interacts for a certain time $t_\mathrm{th}$ with the thermal bath at each step. Let us introduce a parameter $\alpha$, such that for $\alpha=1$, there is no interaction, and for $\alpha=0$ the system thermalizes perfectly in each step. In general, $\alpha=\alpha(t_\mathrm{th})$ is a function of the interaction time with the bath in each step. For $\alpha=0$ the second law of thermodynamics implies that for generic initial states the  extracted work satisfies (see e.g. \cite{Nulton1985,Anders13,Popescu2013b,Reeb2014})
\begin{align}
W = \Delta F - \frac{\Gamma}{N}  +\mathcal{O}\left(\frac{1}{N^2}\right)
\end{align}
where $\Delta F$ is the difference of non-equilibrium free energy between the initial and final state as defined in \eqref{DeltaF} and $\Gamma$ depends on the specific protocol and system of interest. Here we find
\begin{align}
W = \Delta F - \left(\Gamma+\Lambda\frac{\alpha}{1-\alpha}\right)\frac{1}{N}+\mathcal{O}\left(\frac{1}{N^2}\right),
\label{ourResult}
\end{align}
where $\Lambda$ is independent of $\alpha$ and $N$. This implies that the scaling of the dissipation $\Delta F- W$ with $N$ remains the same even with imperfect thermalizations, and it shows  how $N$ can  be rescaled in order to keep the dissipation controlled for different $\alpha$'s.

The result \eqref{ourResult} is derived in two different frameworks. We start by considering a framework where thermalization takes place due to repeated interactions between the system (S) and elements of a thermal bath (B), in the spirit of collision models \cite{Scarani2002,ziman2005description,Filipowicz1986,Caves1987}. We study work extraction from a qubit system S when having access to  B, which consists of a   collection of qubit systems in a thermal state. This allows for a fully inclusive approach, where S,  B and the work storage device are described as quantum systems evolving unitarily. In this case, we find that $\alpha$ in \eqref{ourResult} behaves as 
\begin{align}
\alpha= \cos^2( g t_\mathrm{th}/ \hbar)
\label{alphatuni}
\end{align}
where $g$ is the coupling between S and each qubit of B. For this model, we also compute explicitly $\Gamma$ and $\Lambda$, finding the simple relation
\begin{align}
\Gamma=\frac{\Lambda}{2}, \label{gammalambda}
\end{align} 
for a large class of protocols. 
This result is then generalized to arbitrary qudit systems. Furthermore, we also study  work fluctuations, showing that for a large number of steps the final work distribution using such uncontrolled interactions converges to the same form as for the optimal protocol \cite{Popescu2013b,Aberg2013,AbergWork}.

In a second part of the article, we study the same question in a different and widely used framework, namely when B is a macroscopic bath  and work is extracted by changing the Hamiltonian of $S$ over time. In this case, partial thermalizations naturally appear due to finite-time contacts with B. However, we generalize from the partial thermalizations presented above, by describing the action of B on S as essentially any non-trivial quantum channel with the Gibbs state as a unique fixed point. In this case, we also show \eqref{ourResult}, where $\alpha$ now quantifies how much closer (in trace norm) the state becomes to the corresponding Gibbs state upon application of the channel. Here, one may naturally expect (at least for sufficiently large interaction times)
\begin{align}
\alpha(t_\mathrm{th}) \approx e^{-t_{\mathrm{th}}/\tau_{\mathrm{th}}},
\label{alphatexp}
\end{align}
 where $\tau_{\mathrm{th}}$ is a thermalization time-scale. 

Finally, we study the implications of \eqref{alphatuni} and \eqref{alphatexp} in \eqref{ourResult} when we fix the total time $\mathcal{T}=N t_\mathrm{th}$ of the process and optimize $t_\mathrm{th}$ to minimize the dissipation $W_{\rm dis}\equiv \Delta F- W$ (this definition ensures that $W_{\rm dis}>0$ given our sign convention, i.e., $W>0$ when work is extracted from the system). We find very different behaviours of $W_{\rm dis}$ as a function of the thermalization time $t_{\rm th}$ depending on the model in consideration: while in collision-like models short $t_{\rm th}$ are not desirable, the opposite holds for an exponential relaxation as in \eqref{alphatexp}, which is expected in situations where S is constantly coupled to B. 

The paper is structured as follows. In Section \ref{Framework}, after introducing the framework, we describe the ideal protocol for work extraction and later introduce a protocol which also works for imperfect operations. In Section \ref{ResQ}, we show that the noisy protocol, like the ideal protocol,  allows for maximal work extraction in the case where system and bath consist of qubits, and calculate the corresponding work fluctuations. These results are later generalized in Section \ref{ResG} to generic qudit systems. In section \ref{sec:quenchesandthermalization}, we discuss the second framework. In Sec. \ref{sec:waitingtimes} we discuss how $W_{\rm dis}$ is minimized for the different frameworks, and conclude in Section \ref{Conc}.

\section{Collision Model}
\label{CollisionModel}
Collision models have been extensively used to describe open quantum systems \cite{ziman2005description,Filipowicz1986,Caves1987}, in particular thermalization processes \cite{Scarani2002,Ziman2002,Gennaro2015,cusumano2018entropy,Man2018} and non-equilibrium scenarios in different thermodynamic contexts \cite{diosi2006exact,Raam2014,barra2015thermodynamic,Lorenzo2015,pezzutto2016implications,Strasberg2017}. The main assumption is that when the system interacts with a bath, it collides and thereby interacts with only one particle of the bath at a time. This means that we can model a thermalization process as successive interactions of the system with different parts of the bath.\\
Recently isothermal processes have also been described in the framework of collision models as a concatenation of interactions between the system S and elements of the bath B \cite{Popescu2013,Popescu2013b,Reeb2014}.  A difference between the latter and the standard description of  thermalization  via collision models \cite{Scarani2002} is that in the latter the system S interacts with non-identical Gibbs states, in order to construct an isothermal process.  We follow this approach in the first part of the paper.  We do so in an inclusive way, where all elements entering in the process are described explicitly as quantum systems, in the spirit of the resource theory of thermodynamics \cite{ResourceTheory,Nanomachines,Gour13,mueller2017correlating}.

\subsection{Framework}
\label{Framework}

The model we consider for describing optimal processes consists of a system S, which interacts with a thermal bath B at fixed temperature $T$, and a work battery W, in which the extracted work can be stored. Most of our results concern the scenario in which S is a two-level system, a qubit -- later we will consider generic qudit systems.
 The Hamiltonian of S is hence taken to be $H_{\rm S}=\epsilon_{\rm S} \ket{1}\bra{1}_{\rm S}$. Similarly, the bath consists of a collection of $N+1$ qubits, with internal Hamiltonians $H_{\rm B}^{(k)}= E_k \ket{1}\bra{1}_{\rm B^{(k)}}$, $k=0, ...,N$. For better readability, we will in the following denote any object $X$ acting on $\rm B^{(k)}$ by $X_B^{(k)}$. For W, we consider a weight that stores the work as potential energy and take the Hamiltonian $H_{\rm W}=mg \hat{x}$, where $\hat{x}$ is the continuous position operator, $\hat{x}=\int dx \hspace{0.5mm} x  \hspace{0.5mm}\ketbra{x}{x}$,  and for simplicity we choose $mg=1\ \text{J/m}$. The total Hamiltonian on SBW (which is a shorthand for $S\otimes B \otimes W$) hence reads
 \begin{align}
H= H_{\rm S} + \sum_k H_{\rm B}^{(k)}  + H_{\rm W}.
\end{align}
The initial state is taken to be a pro\-duct state,
 \begin{align}
\rho^{(0)} =\rho_{\rm S}^{(0)} \otimes  \left( \bigotimes_{k=0}^{N} \tau_{\rm B}^{(k)}\right) \otimes \rho_{\rm W}^{(0)}.
\label{initial_state}
\end{align}
We take for S a diagonal state,
\begin{align}
\rho_{\rm S}^{(0)}=(1-p_0)\ket{0}\bra{0}_{\rm S}+p_0\ket{1}\bra{1}_{\rm S}.
\end{align}
The case of coherent states is discussed in \cite{Aberg2013,Korzekwa2016}, where it is shown that the coherent content of the free energy can be extracted by using an external source of coherence, a process where B does not intervene. Hence, here we only consider diagonal states.
 The states of the bath $\tau_{\rm B}^{(k)}$ are Gibbs states at temperature $k_B T=1/\beta$,
\begin{align}
 \tau_{\rm B}^{(k)}=\frac{e^{-\beta H_{\rm B}^{(k)}}}{\Tr[e^{-\beta H_{\rm B}^{(k)}}]}. 
 \end{align}
 Since we are dealing with qubits, we can also write
 \begin{align}
  \tau_{\rm B}^{(k)}=(1-q_k)\ket{0}\bra{0}_{\rm B}+q_k\ket{1}\bra{1}_{\rm B},
 \end{align}
where we note the relation \mbox{$E_k=k_BT\ln \left((1-q_k)/q_k\right)$}. 
Finally, the initial state for W is taken to be $\rho_{\rm W}^{(0)}=\ket{0}\bra{0}_{\rm W}$. As we now argue, this choice is in fact irrelevant for the extracted work.

Thermodynamic operations consist of unitary operations $U$ on SBW. The average extracted work is then defined as the energy gained by W,
 \begin{align}
W := \Tr \left(H_{\rm W}(\rho_{\rm W}'-\rho_{\rm W})\right)
\label{defw}
 \end{align} 
 where $\rho_{\rm W}=\Tr_{\rm SB}  [\rho]$ is the initial state, and $\rho_{\rm W}'=\Tr_{\rm SB} [U \rho U^{\dagger}]$ is the state after the transformation. Importantly, we impose two conditions on  $U$:
 \begin{enumerate}
 \item The total energy must be preserved by $U$, more concretely \cite{ResourceTheory}, \label{cond1}
 \begin{align}
 [U,H]=0.
 \label{energy_conservation}
 \end{align}
 This condition ensures that all energy fluxes are accounted for, as there cannot be  energy flowing out of SBW, and also justifies definition \eqref{defw}.
 \item \label{second}  $U$ must commute with the displacement operator on W. If $\rho_{\rm W}$ is diagonal, this implies that the extracted work $W$ is independent of the initial state $\rho_{\rm W}$ of W. Since we will only consider diagonal states for $\rho_{\rm W}$, this condition ensures that the state of the weight cannot be used as a source of free energy, so that the extracted work comes solely from SB.  
 \end{enumerate}

\begin{figure}[t]
\centering
\includegraphics[width=0.47\textwidth]{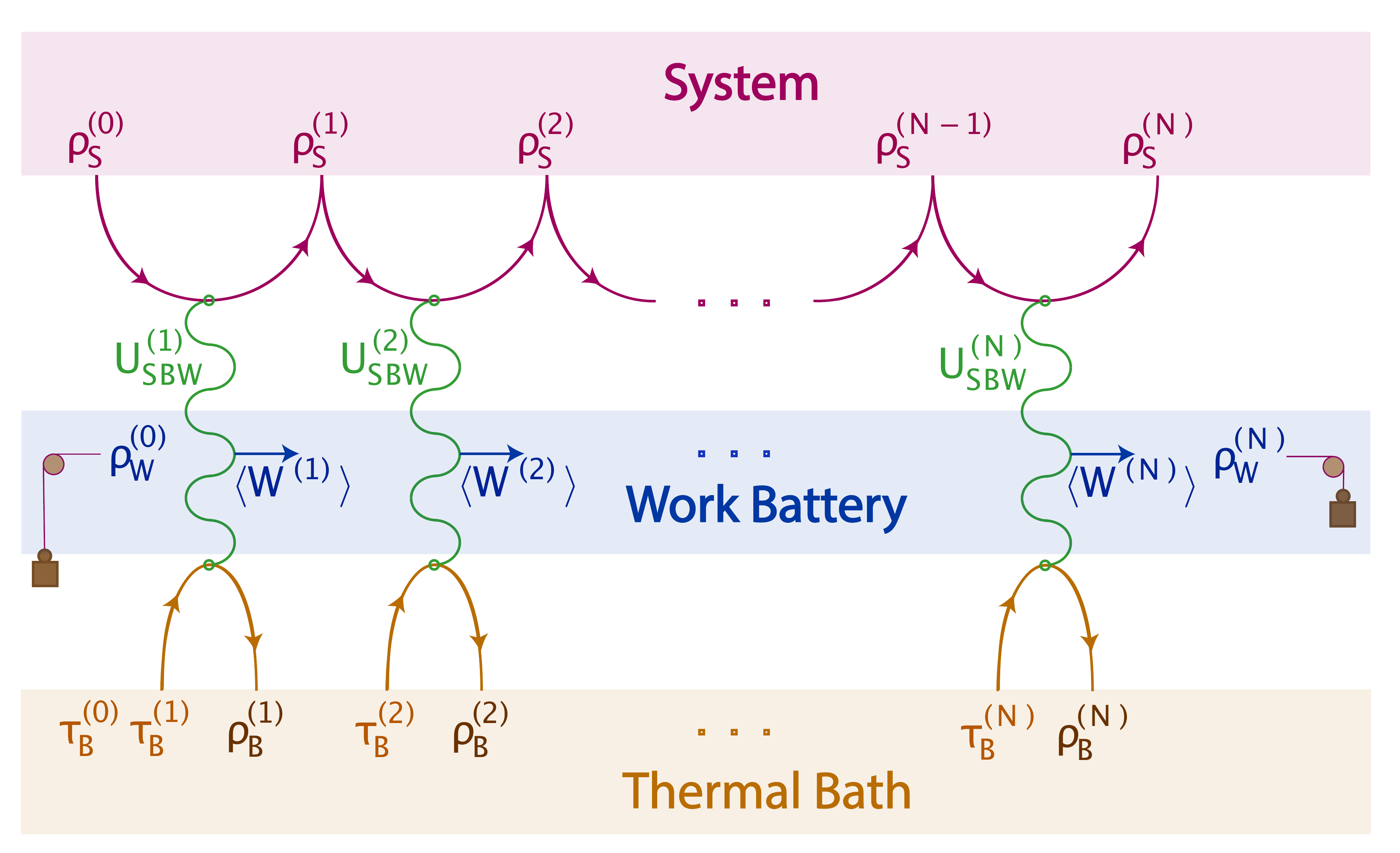}
\caption{
The setting described in this paper consists of three subsystems: The system S initially in a non-thermal state, a work battery W in which the work is stored and a thermal bath B that provides thermal energy to be converted to work. They interact sequentially by means of $U_{\rm SBW}^{(k)}$ such that in each step another part of the thermal bath is involved. The total extracted work can then be computed as the sum of the extracted work in each step, $ W^{(k)} $. 
}
\label{framework}
\end{figure}

\subsubsection{Maximal Work Extraction}

The maximum extractable work of $\rho_{\rm S}^{(0)}$ is given by the change of (non-equilibrium) free energy between initial and final state \cite{Alicki2004,Esposito2011}
\begin{align}
\Delta F \equiv F(\rho_{\rm S}^{(0)},H_{\rm S})-F(\tau_{\rm S},H_{\rm S}),
\label{DeltaF}
\end{align}
where $F(\rho,H) = \Tr (\rho H) - T S(\rho)$, with $S(\rho)$ the von Neumann entropy $S(\rho)=-\Tr [\rho \ln \rho ]$  and 
\begin{align}
\tau_{\rm S} := \frac{e^{-\beta H_{\rm S}}}{\Tr[e^{-\beta H_{\rm S}}]}.
\end{align} 
Note that $\Delta F$ in \eqref{DeltaF} is the difference between initial and final free energy (and not vice versa), which is convenient for our sign convention in which $W>0$ when work is extracted from the system.
 
\begin{figure}[t]
\centering
\includegraphics[width=0.47\textwidth]{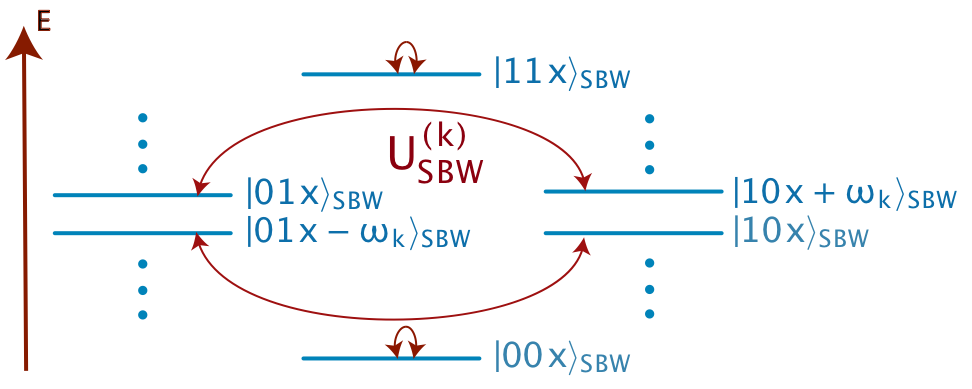}
\caption{
This picture shows the ideal case, where in the $k^{\rm th}$ step the unitary $U^{(k)}_{\rm SBW}$ swaps the energy degenerate states $\ket{0,1,x}_{\rm SBW}$ and $\ket{1,0,x+\omega_k}_{\rm SBW}$ for all $x$, where only the $k^{\rm th}$ bath qubit is involved. The states $\ket{0,0,x}_{\rm SBW}$ and $\ket{1,1,x}_{\rm SBW}$ are mapped to themselves. }
\label{degeneratesubspace}
\end{figure}

Following Refs. \cite{Popescu2013,Popescu2013b}, we now briefly discuss an explicit protocol leading to \eqref{DeltaF}. 
In order to construct an optimal process, the excitation probabilities $\{q_k\}_{k=0}^N$ are chosen to start at $q_0=p_0$ and increase strictly monotic to $q_N=e^{-\beta \epsilon_{\rm S}}/(1+e^{-\beta \epsilon_{\rm S}})$.  Hence, $\tau_{\rm B}^{(0)}=\rho_{\rm S}^{(0)}$ and in the case of maximal work extraction $\tau_{\rm B}^{(N)}=\tau_{\rm S}$. The total process is described by the following unitary operation on SBW,
\begin{align}
U = \prod_{k=1}^{N} U^{(k)}_{\rm SBW}
\label{UT}
\end{align}
where each $U^{(k)}_{\rm SBW}$ acts on S, W and the $k^{\rm th}$ bath qubit, see Figure \ref{framework}. Using the short hand notation, $\ket{a,b,c} := \ket{a}_{\rm S} \otimes \ket{b}_{\rm B} \otimes \ket{c}_{\rm W}$ and $\omega_{k} := E_k-\epsilon_{\rm S}$,  $U^{(k)}_{\rm SBW}$  takes the form
\begin{align}
\begin{split}
U^{(k)}_{\rm SBW}= \int_x & dx \bigg( \ket{0,1,x} \bra{1,0,x+\omega_{k}} +h.c. \\
&+\ket{0,0,x}\bra{0,0,x} +\ket{1,1,x}\bra{1,1,x}\bigg)
\label{U_SBW^k}
\end{split}
\end{align}
Let us also write this expression in matrix form using the basis $\{ \ket{0,0,x}, \ket{0,1,x}, \ket{1,0,x+\omega_k}, \ket{1,1,x}\}$
\begin{align}
U^{(k)}_{\rm SBW}&= \int_x dx \begin{pmatrix}
1 &  0 & 0 & 0 \\
0 & 0 & 1 & 0 \\
0 &  1 & 0 & 0 \\
0 &  0 & 0 & 1
\end{pmatrix}_x,
\end{align}
where the subscript $x$ indicates that the basis depends on $x$.
The action of $U^{(k)}_{\rm SBW}$ is illustrated in Figure \ref{degeneratesubspace}. Note that 
\begin{itemize}
\item $U^{(k)}_{\rm SBW}$ acts upon a degenerate energy subspace, and hence condition \ref{cond1}., as given by  \eqref{energy_conservation}, is satisfied. 
It is straightforward to see that condition \ref{second}. is  satisfied as well. 
\item At the global level, the final state after applying $U^{(k)}_{\rm SBW}$ becomes correlated. At the level of the reduced states, it leads to a swap between the states $\rho_{\rm S} \leftrightarrow \tau^{(k)}_{\rm B}$, 
\item As $k$ increases, the state of S progressively changes from $\rho_{\rm S}^{(0)}$ to $\tau_{\rm S}$, which mimics an isothermal process.
 \end{itemize}
Furthermore, the total extracted work $ W  =\Tr \left(H_{\rm W} (U\rho^{(0)}U^{\dagger}-\rho^{(0)}) \right)$ from \eqref{UT} satisfies,
\begin{align}
 W = \Delta F - \mathcal{O}(1/N),
\label{resultPaul}
\end{align}
 hence becoming maximal in the limit $N\rightarrow \infty$ (see e.g. \cite{Popescu2013,Popescu2013b}).

\subsubsection{Imperfect Operations}

Let us now describe the set of imperfect operations we consider. For clarity, we start by presenting a simple model to describe lack of control on the operations $U_{\rm SBW}^{(k)}$. For that, note that the unitary \eqref{U_SBW^k} can be implemented by turning on the interaction $g\sigma_x$ in the  subspace $\ket{0,1,x}$, $\ket{1,0,x+\omega_x}$ for a time $\pi  \hbar / 2g$. If the timing is not precise, the corresponding  unitary  $\tilde{U}^{(k)}_{\rm SBW}$ takes the form, 
\begin{align}
\tilde{U}^{(k)}_{\rm SBW}&= \int dx \begin{pmatrix}
1 &  0 & 0 & 0 \\
0 & \sqrt{\alpha_k} & i \sqrt{1-\alpha_k} & 0 \\
0 &  i \sqrt{1-\alpha_k} & \sqrt{\alpha_k} & 0 \\
0 &  0 & 0 & 1
\end{pmatrix},
\label{tUi}
\end{align}
where  $\alpha_k\in [0,1]$ quantifies the error in the $k^{\rm th}$ step: for $\alpha_k=0$ we apply the desired full rotation and for $\alpha_k=1$ we do nothing on the state.
After \eqref{tUi}, the reduced states of S and B, given by $\rho_{\rm S/B}^{(k)}=\Tr_{\rm BW/SW} \left( \tilde{U}^{(k)}_{\rm SBW} \rho^{(k)} \tilde{U}^{(k)\dag}_{\rm SBW} \right)$, take the form
\begin{align}
\begin{split}
\rho^{(k)}_{\rm S}&= \alpha_k \rho^{(k-1)}_{\rm S} + (1-\alpha_k) \tau_{\rm B}^{(k)}
\\
\rho_{\rm B}^{(k)} &= \alpha_k \tau_{\rm B}^{(k)} + (1-\alpha_k) \rho^{(k-1)}_{\rm S}.
\label{modelnoiseIII}
\end{split}
\end{align}
Hence, here the lack of control essentially leads to a partial thermalization process. 

In fact, there are more general  processes that lead to \eqref{modelnoiseIII}. In Appendix \ref{appA}, we consider arbitrary thermal operations \cite{ResourceTheory} on the relevant degenerate subspaces $\ket{0,1,x}$, $\ket{1,0,x+\omega_x}$.
By this, we aim to model the situation where the relevant subspace where the experimentalist is acting upon is not thermally isolated, e.g., there is decoherence in the unitary operations being implemented. 
 We then show that all operations of this form lead to the same model of noise \eqref{modelnoiseIII}, with the $\alpha_k$'s being complex functions of the applied specific thermal operation. Therefore, we conclude that \eqref{modelnoiseIII} is generally valid when we restrict ourselves to the operations described above which map states from the degenerate energy subspaces $\ket{0,1,x}$, $\ket{1,0,x+\omega_x}$ to states on that same subspace while acting arbitrarily on an auxiliary system. This is the model we will consider for the rest of our work.
As a remark, note that \eqref{modelnoiseIII} naturally allows for generalizations when considering systems beyond qubits.

Before moving further, it is important to keep in mind that the set of imperfect operations we described is of course not the most general noisy, or thermal, operation.
 The motivation behind our choice is that, as we will now show,  thermodynamic processes can still be constructed despite the lack of control. Conversely, if we lift the restrictions made above, it is not difficult to convince oneself that other types of noise, e.g., thermal noise acting on S or W only, are detrimental for thermodynamic purposes and cannot be compensated for.  This is  discussed in Appendix \ref{appB}.

\subsection{Results}
\label{ResQ}

 We now proceed to derive our main results, taking \eqref{modelnoiseIII} as a starting point. We recall that the $\alpha_k$'s characterize the noise level of the operation in the $k^{th}$ step. One may think of the $\alpha_k$ as being equal to a fixed value $\alpha$. For purpose of generalization, we treat the $\alpha_k$'s as independent random variables with average value $ \alpha$. More precisely, let  the average of a function $f$ over  $\{ \alpha_j \}_{j=1}^{N}$ be
\begin{align}
\langle f(\{\alpha_j \}) \rangle_{\{\alpha_j \}} = \int ... \int \prod_{i=1}^N d\alpha_i P_i(\alpha_i) f(\{\alpha_j \}),
\end{align}
where $P_i(\alpha_i)$ are some unknown probability distributions that satisfy $\int d \alpha_i P_i(\alpha_i) \alpha_i=\alpha$ for all $i$. Since the $\alpha_j$'s are independent, in particular
\begin{align}
\langle \prod_{j=1}^k \alpha_j \rangle_{\{\alpha_j \}} = \alpha^k.
\end{align}
It is then straightforward to obtain from \eqref{modelnoiseIII} that
\begin{align}
\begin{split}
\langle \rho^{(k)}_{\rm S} \rangle_{\{\alpha_j \}_{j=1}^k} &= \alpha \langle\rho^{(k-1)}_{\rm S}\rangle_{\{\alpha_j \}_{j=1}^{k-1}} + (1-\alpha) \tau_{\rm B}^{(k)}
\\
\langle\rho_{\rm B}^{(k)}\rangle_{\{\alpha_j \}_{j=1}^k}  &= \alpha \tau_{\rm B}^{(k)} + (1-\alpha) \langle \rho^{(k-1)}_{\rm S}\rangle_{\{\alpha_j \}_{j=1}^{k-1}}.
\label{modelnoiseII}
\end{split}
\end{align}
where we used that $\tau_{\rm B}^{(k)}$ and $\rho_{\rm S}^{(k-1)}$ are independent of $\alpha_k$. That is, when considering the average over $\{ \alpha_j \}_{j=1}^{N}$, we can just replace $\alpha_k$ by its average $\alpha$. Given that all quantities of interest (average work, fluctuations) are linear functions of $\rho^{(k)}_{SB}$, the same rule follows for their average. In what follows, we hence replace the model  \eqref{modelnoiseIII} by
\begin{align}
\begin{split}
\rho^{(k)}_{\rm S}&= \alpha \rho^{(k-1)}_{\rm S} + (1-\alpha) \tau_{\rm B}^{(k)}
\\
\rho_{\rm B}^{(k)} &= \alpha \tau_{\rm B}^{(k)} + (1-\alpha) \rho^{(k-1)}_{\rm S},
\label{modelnoise}
\end{split}
\end{align}
where the corresponding results should be understood as the average over uncontrolled parameters $\{ \alpha_j\}$.

Now, starting from the recursive definition of $\rho^{(k)}_{\rm S}$ in \eqref{modelnoise}, we can determine its explicit form as
\begin{align}
\label{recrel}
	\rho_{\rm S}^{(k)}=(1-\alpha) \sum_{i=1}^{k}\alpha^{k-i} \tau^{(i)}_{\rm B}+\alpha^k \rho_{\rm S}^{(0)}.
\end{align}
Furthermore, all $\tau_{\rm B}^{(k)}$ are diagonal, and as we assumed that $\rho_{\rm S}^{(0)}$ is also diagonal,  so is $\rho_{\rm S}^{(k)}$, and hence the same relation holds for the excitation probabilities $p_k$, with $\rho_{\rm S}^{(k)}=(1-p_k) \ketbra{0}{0}_{\rm S}+p_k \ketbra{1}{1}_{\rm S}$,
\begin{align}
	p_k=(1-\alpha) \sum_{i=1}^{k}\alpha^{k-i} q_i+\alpha^k p_0.
\label{explicit}
\end{align}

\subsubsection{Average Work Extraction}
\label{sec3a}
We first show that it is possible to obtain $W \rightarrow \Delta F$ for any $\alpha<1$ in the limit of $N \rightarrow \infty$. For clarity of the presentation, we find it convenient to  present here the case where  $\rho_{\rm S}=\ketbra{0}{0}_{\rm S}$ and $\epsilon_{\rm S}=0$ and discuss the general case for arbitrary initial states of S and higher dimensional systems in Section~\ref{ResG} and Appendix~\ref{appGen}. Note that the case  $\rho_{\rm S}=\ketbra{0}{0}_{\rm S}$ and $\epsilon_{\rm S}=0$ corresponds to  well-known  information to energy conversion, where $\Delta F =k_B T \ln 2$ work can be extracted from one bit of information. The opposite direction, resetting one bit of information while investing work $\Delta F =k_B T \ln 2$, the so-called erasure, works analogously since we are looking at reversible processes.

\begin{figure}[t!]
\centering
\includegraphics[width=0.47\textwidth]{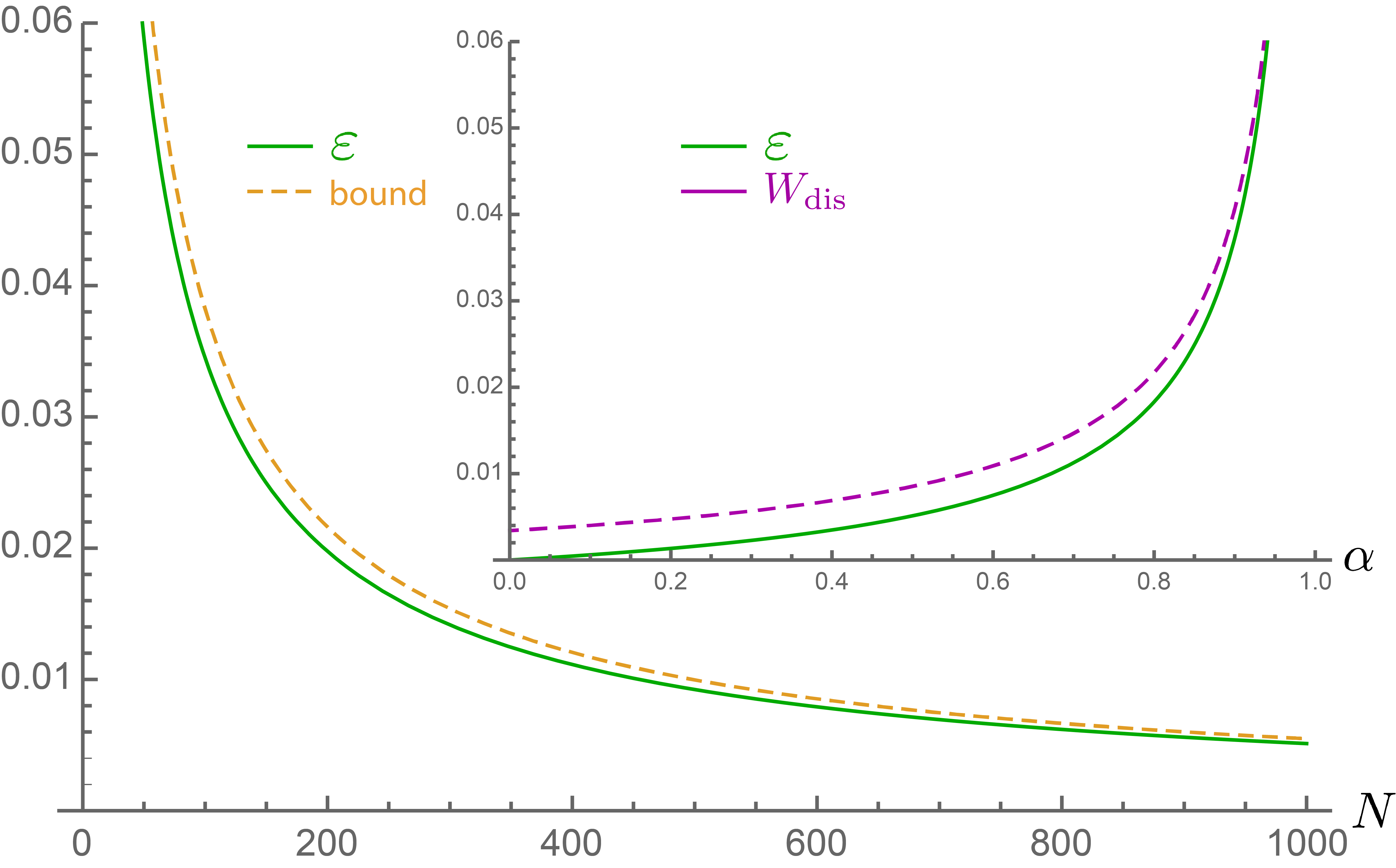}
\caption{{\bf Main figure:} The loss $ \varepsilon $ due to noise as a function of the total number of steps $N$ for $\alpha=1/2$, $p_k= k/2N$, $k_BT\ln 2=1$.  The green line corresponds to the exact loss $ \varepsilon $ computed numerically with $N$ steps, whereas the orange (dashed) line corresponds to the upper bound  \eqref{upperbound}.  {\bf Inset figure:} Errors as a function of $\alpha$ for  $N=1000$, $p_k=k/2N$, $k_BT\ln 2=1$. The green line corresponds to $ \varepsilon $, while the purple (dashed) one is the total dissipation $W_{\rm dis} $.
}
\label{errorbound}
\end{figure}

To compute the average energy gain of the weight W, we first notice that the work extracted in step $k$ is
\begin{align}
W^{(k)} &= \Tr\hspace{-0.5mm} \left[  \big(\rho_{\rm W}^{(k)} - \rho_{\rm W}^{(k-1)} \big) H_{\rm W}\right]
\nonumber \\
&=\Tr \hspace{-0.5mm} \left[ H_{\rm B}^{(k)} \!\left( \tau_{\rm B}^{(k)} \!-\!\rho_{\rm B}^{(k)}  \right) \!-\! H_{\rm S}\! \left( \rho_{\rm S}^{(k)} \!-\! \rho_{\rm S}^{(k-1)}  \right) \right] \!, 
\end{align}
which follows from energy conservation. Thus, the total average work extracted after $N$ steps is the sum
\begin{align}
 W &= \sum_{k=1}^{N}\! \Tr \! \left[ H_{\rm B}^{(k)} \!\left( \tau_{\rm B}^{(k)} \!-\! \rho_{\rm B}^{(k)}  \right) \!-\! H_{\rm S} \!\left( \rho_{\rm S}^{(k)} \!-\! \rho_{\rm S}^{(k-1)}  \right) \right] \nonumber \\ 
 &= (1\!-\!\alpha) \sum_{k=1}^{N} \Tr \left[ \left( H_{\rm B}^{(k)}\! -\! H_{\rm S}  \right)  \left( \tau_{\rm B}^{(k)}\! - \! \rho_{\rm S}^{(k-1)} \right) \right]  \label{generalwork}\!.
\end{align}
For the qubit case, we have 
\begin{align}
 W &= (1-\alpha) \sum_{k=1}^N (E_k-\epsilon_{\rm S}) ( q_k-p_{k-1} ).
\end{align}
Let us now use our choice of initial conditions $\epsilon_{\rm S}=0$ and $p_0=0$, i.e., initially  $\rho_{\rm S}=\ket{0}\bra{0}_{\rm S}$. The distribution of bath qubits is taken to be
\begin{align}
q_k=\frac{k}{2N} \quad \Rightarrow \quad \Delta q_k:=q_k-q_{k-1} = \frac{1}{2N}\ \ \forall k,
\label{q_k}
\end{align}
so that $\tau_{\rm B}^{(0)} = \rho_{\rm S}$ and $\tau_{\rm B}^{(N)} = \tau_{\rm S}$. Note that this specific distribution is chosen here for simplicity and to give an explicit upper bound -- we later discuss more general settings. Using \eqref{explicit}, we get an explicit form for the excitation probabilities of $\rho_{\rm S}^{(k)}$,
\begin{align}
p_k&=(1-\alpha)\sum_{i=1}^k \alpha^{k-i}\frac{i}{2N}=q_{k}-\frac{\alpha (1-\alpha^k)}{1-\alpha}\Delta q_k,
\label{excprob}
\end{align}
where we can already see that they only differ by $\mathcal{O}\left(1/N\right)$ compared to the excitation probabilities $q_k$ we would get using error-free unitaries. That is, for large $N$ our state $\rho_S$ is at every step  close to the corresponding thermal state, and hence it follows approximately the isothermal trajectory of the state transformation. We now show that for large $N$ this also happens at an optimal work cost, i.e., that the work extracted becomes  arbitrary close to the one in the error-free case. Using \eqref{excprob}, the average extracted work simplifies to
\begin{align}
 W  &= (1-\alpha) \sum_{k=1}^{N}  E_k  \left( q_k - p_{k-1} \right)\nonumber\\
&=(1-\alpha) \sum_{k=1}^N E_k \left(q_k - q_{k-1}+\frac{\alpha-\alpha^k}{1-\alpha} \Delta q_k  \right)\nonumber\\
&=\sum_{k=1}^N E_k \Delta q_k (1-\alpha^{k}) \nonumber \\
&=\Delta F - \frac{\Gamma}{N} - \varepsilon - \ord{\frac{1}{N^2}} ,
\end{align}
where $\Delta F - \frac{\Gamma}{N} - \ord{\frac{1}{N^2}}$ with $\Gamma = \ord{E_{\rm max}}$ is the maximum average work that can be extracted with $N$ steps in the error-free protocol, i.e., with $\alpha=0$. The loss due to the noise characterized by $\alpha$ is described by $ \varepsilon = \frac{\alpha}{1-\alpha} \frac{\Lambda}{N} + \ord{\frac{1}{N^2}}$ and corresponds to the difference between the work extracted in the optimal protocol ($\alpha=0$) and the one with faulty unitaries $(\alpha\neq 0$). It can be upper bounded by
\begin{align}
 \varepsilon &= \frac{1}{2N}\sum_{k=1}^N E_k \alpha^k
=\frac{k_BT}{2N} \sum_{k=1}^N \ln\left( \frac{2N-k}{k} \right) \alpha^k \nonumber\\
& < \frac{k_BT}{2N} \sum_{k=1}^N \ln \left(2N \right) \alpha^k< \frac{\alpha}{1-\alpha} \frac{k_BT\ln (2N)}{2N},
\label{upperbound}
\end{align}
which implies $\Lambda = \ord{E_{\rm max}}$.
 In the limit $N\rightarrow \infty$,  we get $ \varepsilon  \rightarrow 0$ and hence $ W \rightarrow \Delta F$ for any $\alpha <1$. 
In Figure \ref{errorbound} we plot the numerical value of  loss $ \varepsilon $ and the upper bound \eqref{upperbound}. We observe that the deviation is small and the bound almost tight.
We also plot $ \varepsilon $ together with the total dissipation $W_{ \rm dis}=\Delta F - W$ for finitely many steps as a function of $\alpha$. 

\subsubsection{Fluctuations}
\label{secB}

The fluctuations of work are captured by the probability distribution $P(w)=\bra{w} \rho_{\rm W} \ket{w}$, where $\rho_{\rm W} $ is the final state of W, $\rho_{\rm W} = \Tr_{\rm SB} U \rho^{(0)} U^{\dagger}$ and $\rho^{(0)}$ is defined in \eqref{initial_state}. That is, the work fluctuations are mapped to  energy fluctuations of the weight, which initially starts in a well-defined energy \footnote{Note that, since we consider  non-coherent states of S,  the same fluctuations would be obtained by using the more standard two-energy-measurement scheme on SB together \cite{Talkner07}. The two approaches are equivalent here. }.

\begin{figure}[t!]
\centering
\includegraphics[width=0.23\textwidth]{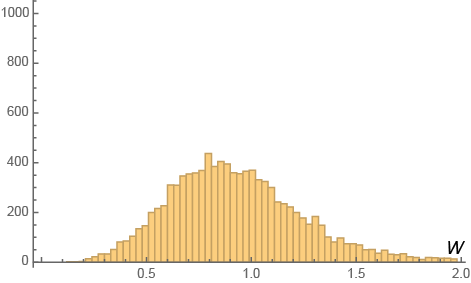}
\includegraphics[width=0.23\textwidth]{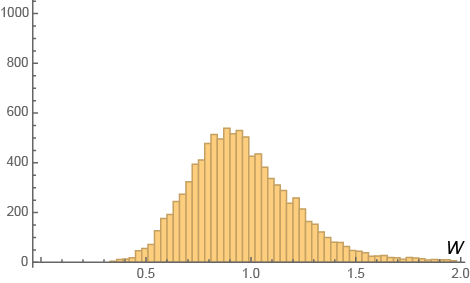}
\scriptsize{$N=100$, $\avw =0.939$, $\sigma= 0.34$ \quad
$N=200$, $\avw=0.967$, $\sigma=0.27$} \\
\includegraphics[width=0.23\textwidth]{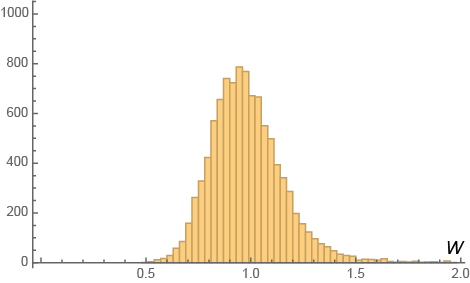}
\includegraphics[width=0.23\textwidth]{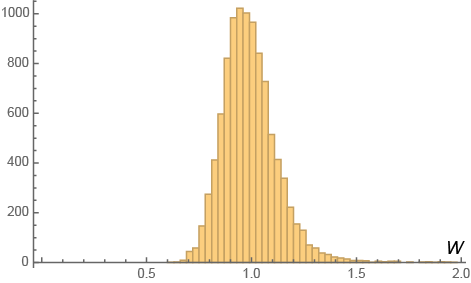}
\scriptsize{$N=500$, $\avw = 0.985$, $\sigma=0.18$ \quad
 $N=1000$, $\avw = 0.993$, $\sigma= 0.14$} \\
\caption{Histograms of the extracted work for $\rho_{\rm S}=\ketbra{0}{0}_{\rm S}$, $\epsilon_{\rm S}=0$, $\alpha=1/2$ and with $k_BT\ln 2=1$. The fluctuations were sampled over 10000 runs to obtain each plot. In agreement with our results, the average work tends to $k_BT\ln 2$ and the fluctuations decrease with increasing $N$.
}
\label{fluctuations}
\end{figure}

 In Appendix \ref{appFluct}, a lengthy calculation shows that for $\rho_{\rm S}=\ketbra{0}{0}_{\rm S}$, $\epsilon_{\rm S}=0$, $\alpha_k\leq \alpha <1 \ \forall k$ and distribution \eqref{q_k}, the variance of $P(w)$ decreases as $\ord{\ln( N) / N}$, and hence disappears in the limit $N\rightarrow \infty$. This result implies that an amount $k_B T \ln 2$ of work can be extracted with \emph{certainty} from one bit of information, even when dealing with partial thermalizations as in \eqref{modelnoise}.   The work probability distribution of this process is illustrated in Figure \ref{fluctuations}.

 Given this result, the case of $\rho_{\rm S} = (1-p_0) \ketbra{0}{0}_{\rm S} +p_0 \ketbra{1}{1}_{\rm S}$ and $\epsilon_{\rm S} \neq 0$ can be easily understood qualitatively. First of all, recall that for general $\rho_{\rm S}$ the optimal protocol  is such that $H_{\rm B}^{(0)}$ satisfies $\tau_{\rm B}^{(0)}=\rho_{\rm S} $ and $H_{\rm B}^{(N)}=H_{\rm S}$. Secondly, since $P(w)$ is linear in $\rho_{\rm S}$, we can write $P(w)= (1-p_0 )P_0 (w) + p_0 P_1(w)$ where $P_{0,1}(w) $ is the probability distribution starting the protocol with  $\ketbra{0}{0}_{\rm S}$ and $\ketbra{1}{1}_{\rm S}$, respectively. From our previous result, it follows that $P_0 (w)$ tends to a peaked distribution, which is now centered at $w_0=F(\rho_{\rm S},H_{\rm B}^{(0)})-F(\tau_{\rm S},H_{\rm S})$ (recall that we consider the protocol for maximal work extraction from $\rho_{\rm S}$). Similarly,  $P_1 (w)$ tends also to a peaked distribution  centered at $w_1=\epsilon_{\rm S} - E_{0} +F(\rho_{\rm S},H_{\rm B}^{(0)})-F(\tau_{\rm S},H_{\rm S})$. Putting everything together, and noting that $w_0 =k_BT \ln((1-p_0)/ (1-p_{\rm S}^{(eq)})) $ and $w_1 =k_BT \ln(p_0/ p_{\rm S}^{(eq)})$, we can write the work probability distribution in the limit $N \rightarrow \infty$ as
 \begin{align}
 P(w)& \stackrel{N\rightarrow \infty}{\longrightarrow} p_0 \delta\left(w -k_BT \ln\frac{p_0}{\ p_{\rm S}^{(eq)}} \right)
 \nonumber\\& +(1-p_0) \delta\hspace{-1mm}\left(w -k_BT \ln\frac{1-p_0}{\ 1-p_{\rm S}^{(eq)} } \right)
 \end{align}
 where $\delta(w-x)$ is the Delta function. Crucially, this distribution is independent of $\alpha$, and hence we recover previous results where perfect thermalization operations ($\alpha=0$) were assumed \cite{Popescu2013b,AbergWork}.

\subsubsection{Generalizations and Discussion}
\label{ResG}

Although our results have been derived for qubit systems, they can be naturally extended to qudits. For that we consider the model of partial thermalizations \eqref{modelnoise}. This implies for $\rho_{\rm S}^{(k)}$ in \eqref{recrel} that (assuming $\rho_{\rm S}^{(0)}=\tau_{\rm B}^{(0)}$)
\begin{align}
\rho_{\rm S}^{(k)}&=(1-\alpha) \sum_{j=0}^{k}\alpha^{j} \tau^{(k-j)}_{\rm B}+\alpha^k \tau_{\rm B}^{(0)} \nonumber\\
&= \tau_{\rm B}^{(k)} \!-\!(1\!-\!\alpha) \sum_{j=0}^{k}\alpha^{j}(\tau^{(k)}_{\rm B} \!-\!\tau^{(k-j)}_{\rm B})\!+\!\mathcal{O}(\alpha^k).
\label{exprhos}
\end{align}
Let us define the continuously differentiable family of Hamiltonians $H(s)$ for $s \in [0,1]$, such that we can draw $H_{\rm B}^{(k)} :=H(k/N)$, and similarly the family of 
states $\tau(s)$ for $s \in [0,1]$, such that $\tau_B^{(j)}=\tau(j/N)$ and $\dot{\tau}(s) \equiv \frac{d}{d s}\tau(s)$. We can then expand $\tau^{(k)}_{\rm B}-\tau^{(k-j)}_{\rm B}$ over $1/N$ in \eqref{exprhos} to obtain,
\begin{align}
\rho_{\rm S}^{(k)}&=\tau_{\rm B}^{(k)}-(1-\alpha) \sum_{j=0}^{k}\frac{j}{N} \alpha^j \dot{\tau}(k/N)+\mathcal{O}(1/N^2)
\nonumber\\
&\approx \tau_{\rm B}^{(k)}-\frac{\alpha}{N(1-\alpha)}\dot{\tau}(k/N)+\mathcal{O}(1/N^2)
\label{expansionrhosk}
\end{align}
Hence, we see how the factor $\alpha/(1-\alpha)N$ in \eqref{ourResult} naturally arises under imperfect thermalizations. It is a consequence of the state $\rho_S^{(k)}$ lagging behind the thermal state $\tau^{(k)}_{\rm B}$. 

In general, computing the average extracted work   is more involved as it depends both on $\tau_{\rm B}^{(k)}$ and the Hamiltonians $H_{\rm B}^{(k)}$. This is done in Appendix \ref{appGen} by  using the continuous expansion described above. In particular, it is convenient to introduce the continuous limit of the non-equilibrium free energy function,
\begin{align}
F(s):=F(\tau(s), H_{\rm S})
\end{align}
 with $s \in [0,1]$.
In Appendix \ref{appGen}, we show (in the case of a full-rank initial state) that $W_{\rm dis}\equiv \Delta F - W $ satisfies at order $1/N$
\begin{align}
W_{\rm dis} =& -\frac{1}{2N}\left(1+\frac{2\alpha}{1-\alpha}\right) \int_0^1 \hspace{-1.5mm}ds \hspace{1mm} \Tr(\dot{H}(s) \dot{\tau}(s))
\nonumber\\&+\frac{\alpha}{N(1-\alpha)}\left(\dot{F}(1)- 2\dot{F}(0_+)\right).
\label{genwork}
\end{align}
The minus sign in the first term arises, since
\begin{align}
\int_0^1 \hspace{-1.5mm}ds \hspace{1mm} \Tr(\dot{H}(s) \dot{\tau}(s)) \leq 0
\end{align}
for any trajectory of Hamiltonians $H(s)$, which, for example, can be seen using the techniques of Refs.~\cite{Perarnau2017,scandi2018thermodynamic}.
The second contribution to the dissipation takes place only at the extremes of the trajectory, when S is put in contact with B for the first/last time. In particular, it is worth noticing that if  $\tau(1)=e^{-\beta H_{\rm S}}/\mathcal{Z}_S$, then we have that $\dot{F}(1)=0$ because $F(s)$ is minimized when the state is thermal. By taking an appropriate trajectory, we can also guarantee $\dot{F}(0)=0$. The more interesting contribution is the first term, which depends on the whole trajectory $H(s)$. Crucially, the dependence on $\alpha$ of the error is independent of the trajectory and is always given by an overall factor $\alpha/(1-\alpha)$. An important consequence of \eqref{genwork} is that protocols that minimize dissipation for $\alpha=0$ (see \cite{scandi2018thermodynamic} for examples) will also minimize dissipation for imperfect thermalizations. 

In the above discussion we assumed $\rho_{\rm S}^{(0)}=\tau_{\rm B}^{(0)}$. This is always possible to fulfill if $\rho_{\rm S}^{(0)}$ has full rank. If this condition is not fulfilled, one obtains an additional error term in the dissipated work, which is detailled in Appendix~\ref{appGen}. 
Let us here discuss the limit of optimal protocols. For any $\delta>0$ it is possible to choose $H(s)$ such that $\norm{\tau_{\rm B}^{(0)} - \rho_{\rm S}^{(0)} } \leq \delta \ll 1$. In this case, one finds that the additional error term scales as $\ord{\delta \log(1/\delta)}$ and similarly 
\begin{align}
- \int_0^1 \hspace{-1.5mm}ds \hspace{1mm} \Tr(\dot{H}(s) \dot{\tau}(s)) = \ord{\log(1/\delta)}.
\end{align}
If we now choose $\delta = \ord{1/N}$ we thus recover the scaling behaviour
\begin{align}
	W_{\rm dis} = \ord{\frac{\log N}{N}},
\end{align}
in accordance with our explicit calculation for the qubit example.

Let us also briefly comment on our assumption that the $\alpha_k$ in \eqref{modelnoiseIII} are random variables with identical expectation value $\alpha$, which is independent of $N$. 
This assumption is in fact not crucial for the main results --although it is very convenient for the computations--, the reason being that our result holds for \emph{any} $\alpha < 1$. Note that we can always decrease the individual values $\alpha_k$ without compromising the work extraction protocol. As a consequence, even if the expectation value of the $\alpha_k$ would change with $k$, one can guarantee convergence to an optimal thermodynamic process as long as $\alpha_k \leq \alpha < 1, \forall k $. Furthermore, even if $\alpha = 1-C/N$ for some constant $C>0$, one still obtains the optimal expected work in the limit $N\rightarrow \infty$ (also see section~\ref{sec:quenchesandthermalization}).  
Similarly, we strongly expect that our results also hold for \emph{typical} realizations of the $\alpha_k$, provided that each $\alpha_k$ is chosen independently with expectation values bounded away from $1$. Indeed, we expect that our results still hold even if a diverging number of $\alpha_k$ is equal to $1$ provided that these $\alpha_k$ are a vanishing fraction of all the $\alpha_k$ in the limit $N\rightarrow \infty$.

It is worth pointing out that even when dealing with qubits, there exist uncontrolled interactions that are thermodynamically detrimental and cannot be compensated for. For example, letting S simply thermalize with B without raising the energy in W can only decrease the extractable work. Similarly, if B and W exchange energy, W may lose energy to the bath. In order to build an optimal thermodynamic process, it is needed that B continuously thermalizes S (and not W) while W is extracting work from it. A more detailed description of interactions that cannot be tolerated can be found in Appendix \ref{appB}.

Finally, it is important to realize that, although we have focused our attention on optimal work extraction, our results also allow for optimal state transformations in the other direction, i.e., they can be used for converting a thermal state to a given final state at optimal work cost. More generally, they apply to the interconversion of any two  states at minimal work cost.  Furthermore, we note that although our results have been derived for a particular model within the resource theory of thermodynamics, we expect that similar results can be obtained in other frameworks, such as \cite{Esposito2010,Reeb2014,Anders13,AbergWork,Gallego14}. One such example is illustrated in the next section.

\section{Time-dependent Hamiltonians and Thermalization Maps}
\label{sec:quenchesandthermalization}
The collision model presented above has the advantage of being very explicit, allowing us to derive precise results both about average work extraction and fluctuations. However, it has the drawback of requiring a fine-tuned bath. We therefore now show similar results in a different framework, in which work is extracted by changing the Hamiltonian of
S over time and contacts with the bath are described by a thermalizing channel, which can essentially be any channel that brings the system closer to the thermal state. The purpose of this section is two-fold (i) this model is experimentally more relevant and realistic, as no control over B is assumed (in the spirit of open quantum systems), and (ii) in this model we are able to generalize our results to arbitrary interactions with the bath that bring the system closer to the thermal state, with partial thermalizations as in the previous section being a specific instance. 

\subsection{Framework}
\label{section:gibbs}
In this framework, a protocol consists of a choice of a trajectory of Hamiltonians $H(t)$ and a choice of $N$ times $\{t_i\}_{i=1}^N$ at which the system S is brought in contact with the bath B (the continuum limit, where $(t_i - t_{i+1})\rightarrow 0$ so that S is essentially always in contact with B, will be discussed below). For simplicity, in the following we only consider cyclic protocols, for which the initial Hamiltonian coincides with the final Hamiltonian and denote by $t_0$ the time at which the protocol starts. We further assume that while the system is in contact with the bath, the Hamiltonian is not changed and the system thus (partially) thermalizes to the Hamiltonian $H^{(i)} = H(t_i)$. 
The contact to the bath at time $t_i$ is effectively described by a quantum channel $G^{(i)}$ and results in a state
\begin{align}
	\sigma^{(i)} = G^{(i)}(\rho^{(i)}),
\end{align}
where $\rho^{(i)}$ is the state of S right before the thermal contact.
Inbetween the contacts to the bath, the system evolves according to the time-dependent Hamiltonian $H(t)$ from time $t_i$ to time $t_{i+1}$, represented by the unitary $U_{t_{i+1},t_i}$. 
The states $\rho^{(i)}$ and $\sigma^{(i-1)}$ are therefore related by the time-dependent evolution under $H(t)$ as
\begin{align}
	\rho^{(i)} = U_{t_i, t_{i-1}} \sigma^{(i-1)} U_{t_i,t_{i-1}}^\dagger.
\end{align}
This time-evolution is associated with a work-gain
\begin{align}
	W^{(i)} = \Tr\left( H^{(i-1)} \sigma^{(i-1)} - H^{(i)}\rho^{(i)} \right),
\end{align}
with $\sigma^{(0)}=\rho^{(0)}$ being the initial state.
Given an initial state $\rho^{(0)}$ and trajectory of Hamiltonians $H(t)$, we denote the total work in an $N$-step protocol by
\begin{align}
	W(\rho^{(0)},H(t)) = \sum_{i=1}^N W^{(i)}. 
	\label{totalW}
\end{align}
Let us finally discuss the assumption that we make on the thermalizing maps $G^{(i)}$. Clearly, the thermalizing maps should have the thermal state $\tau^{(i)}$ of the Hamiltonian $H^{(i)}$ as a fixed-point. 
It then follows in general from the monotonicity of the trace-distance under quantum channels that 
\begin{align}
	\norm{G^{(i)}(\rho) - \tau^{(i)}}&= \norm{G^{(i)}(\rho) - G^{(i)}(\tau^{(i)})}\nonumber
	\\&\leq\norm{\rho- \tau^{(i)}}. 
	\label{generalcond}
\end{align}
In the following we want to model that the thermalizing maps have \emph{some} non-trivial effect on any non-thermal initial state. We  will therefore assume that there exists a fixed number $\alpha < 1$ such that
\begin{align}
	\norm{G^{(i)}(\rho) -\tau^{(i)}} \leq \alpha \norm{\rho-\tau^{(i)}}. 
\end{align}
This expresses formally the physical requirement that the interaction with the bath brings the system closer to the thermal state by at least some, even arbitrarily small, amount. In the following we call $\alpha$ the \emph{thermalization parameter} of the protocol.

Before stating our results, let us emphasize that even though we model each thermalization process using a single map $G^{(i)}$, this does not mean that the physical time $t_{\rm th}^{(i)}$ the thermalization step takes is zero. It merely means that we assume that during the contact with the heat bath the Hamiltonian is constant and the overall effect of the thermalization is effectively modelled by the quantum channel $G^{(i)}$. In particular, our set-up includes the commonly studied case where the changes of the Hamiltonians between the thermalization steps is instantaneous, so-called \emph{quenches}, and the total time of the protocol is simply given by $\mathcal T=\sum_i t_{\rm th}^{(i)}$.

\subsection{Results}\label{resultsgibbsmaps}
Having introduced the framework, we can now easily state our result.
Let us mention before, though, 
that the maximum amount of work that can be extracted using a cyclic protocol of the above form is bounded as \cite{AbergWork,Wilming2016}
\begin{align}
	W(\rho^{(0)},H(t)) &\leq F(\rho^{(0)},H^{(0)}) - F(\tau^{(0)},H^{(0)}) \nonumber\\
	&=\Delta F.
\end{align}
Furthermore, it has been known that for perfectly thermalizing maps ($\alpha=0$), this bound can be achieved to arbitrary accuracy \cite{Anders13,AbergWork,Gallego14,Jacobs2009}. The optimal protocol consists of a first rapid change of the Hamiltonian to a Hamiltonian $\tilde H^{(0)}$ such that (to arbitrary accuracy) 
\begin{align}
	\rho^{(0)} = \frac{\e^{-\beta \tilde H^{(0)}}}{\mathcal{Z}},
\end{align}
followed by an \emph{isothermal process} back to the initial Hamiltonian, meaning a process where $t_i - t_{i-1}=1/N$ and $N\rightarrow \infty$, so that the system stays in the thermal state of the current Hamiltonian throughout the protocol. 
The first quench yields a work-gain 
\begin{align}
	W_{\mathrm{quench}} &= \Tr\left(\rho^{(0)}\left(H^{(0)}-\tilde H^{(0)}\right)\right) \nonumber\\
	&= F(\rho^{(0)},H^{(0)}) - F(\rho^{(0)},\tilde H^{(0)}).
\end{align}
The optimality of the above protocol follows from the fact that an optimal isothermal process between two Hamiltonians $H(0)$ and $H(1)$ with thermal states $\tau(0)$ and $\tau(1)$ has work-gain
\begin{align}
	\Delta F_{\mathrm{iso}}:=F(\tau(0),H(0)) - F(\tau(1),H(1)).
\end{align}
Therefore, the total work-gain in the optimal protocol is given by (using $H(0)=\tilde H^{(0)}$, $H(1)=H^{(0)}$):
\begin{align}
	W_{\mathrm{optimal}} &= W_{\mathrm{quench}} + \Delta F_{\mathrm{iso}} = \Delta F.
\end{align}

\subsubsection{Main Result}
Our following result shows that the protocol discussed above remains optimal for any thermalizing parameter $\alpha<1$, since isothermal processes remain optimal in the limit $N\rightarrow \infty$:
In the above formulation, consider a protocol with a smooth curve of Hamiltonians $H(t)$, thermalization parameter $\alpha<1$ and initial state $\rho^{(0)}$. Choose the time-steps as $t_i=i/N$. Then the dissipated work can be expressed (to leading order in $1/N$) as 
	\begin{align}
		W_{\rm dis} &= \Delta F_{\mathrm{iso}} - W_{\mathrm{iso}}(\rho^{(0)},H(t)) \nonumber \\
		&= -\frac{1}{2N}\int_0^1\! ds \Tr\left(\dot \tau(s) \dot H(s)\right) \!+\! \frac{\alpha}{1-\alpha}\frac{\Lambda}{N} \!+\! \frac{K}{N}. \label{eq:averageiso}
	\end{align}
The dependence on the initial state $\rho^{(0)}$ only appears in the constant $K$. 
The (relatively involved) proof of \eqref{eq:averageiso} is given in Appendix~\ref{appframe}. 
Importantly, we have $K=0$ in the case where we assume instantaneous quenches instead of continuous unitary time-evolution between the thermalization steps. Furthermore, a calculation similar to that provided in Appendix \ref{appGen}, shows that for the case of instantenous quenches and partial thermalizations as in \eqref{modelnoiseIII}, one recovers the same result as \eqref{genwork} with $\dot{F}(1)=0$ and $\dot{F}(0_+)=0$, making evident the connection between both frameworks.

\subsubsection{Fluctuations}

So far, we have only discussed average work extraction. This has a good reason: In the current framework it is in general not possible to associate a classical random variable that measures the fluctuating work to the process due to the coherences that can be created by the unitaries $U_{t_i,t_{i-1}}$ and the channels $G^{(i)}$ \cite{Perarnau-Llobet2016}.
Let us emphasize, however, that in the limit $N\rightarrow \infty$ in which the extracted work becomes optimal, these coherences become arbitrarily small. It then follows that we can associate a distribution of fluctuating work to the limiting process, which coincides with the limiting process when perfect thermalization and quenches of the Hamiltonian are assumed. 
 It follows that this limiting distribution is universal, i.e., independent of the trajectory $H(t)$ and coincides with the one obtained from the collision model, see e.g.  \cite{AbergWork}.\\

\section{Optimal Thermalization Times to Minimize Dissipation}
\label{sec:waitingtimes}

In this section, we study how the interaction time with the bath in each step affects the total dissipated work when we fix the total time $\mathcal{T}$ of the protocol. We focus on \eqref{genwork} with $\dot{F}(1)=\dot{F}(0_+)=0$, i.e. the case where the dissipation satisfies,
\begin{align}
W_{\rm dis} = 2\,\Gamma \left(\frac{1}{2}+\frac{\alpha(t_\mathrm{th})}{1-\alpha(t_\mathrm{th})}\right)\frac{t_{\mathrm{th}}}{\mathcal{T}},
\label{Wdissopt}
\end{align}
provided that $N=\mathcal{T}/t_{\mathrm{th}}$ is sufficiently large, and with $\Gamma = -\frac{1}{2} \int_0^1 ds \Tr\left[\dot \tau(s) \dot H(s)\right] $. Since the dissipation is proportional to $\Gamma$ and $\mathcal{T}^{-1}$, we can focus on the function,
\begin{align}
G(t_{\mathrm{th}})= \left(\frac{1}{2}+\frac{\alpha(t_\mathrm{th})}{1-\alpha(t_\mathrm{th})}\right)t_{\mathrm{th}}.
\end{align}
Minimizing $G(t_{\mathrm{th}})$ will provide the optimal thermalization time $t_{\mathrm{th}}$ that minimizes dissipation. In fact, our goal is to study how  $G(t_{\mathrm{th}})$ behaves depending on the form of equilibration between system and bath, which is encapsulated in $\alpha(t_\mathrm{th})$. 

\begin{figure}[t!]
\centering
\includegraphics[width=0.47\textwidth]{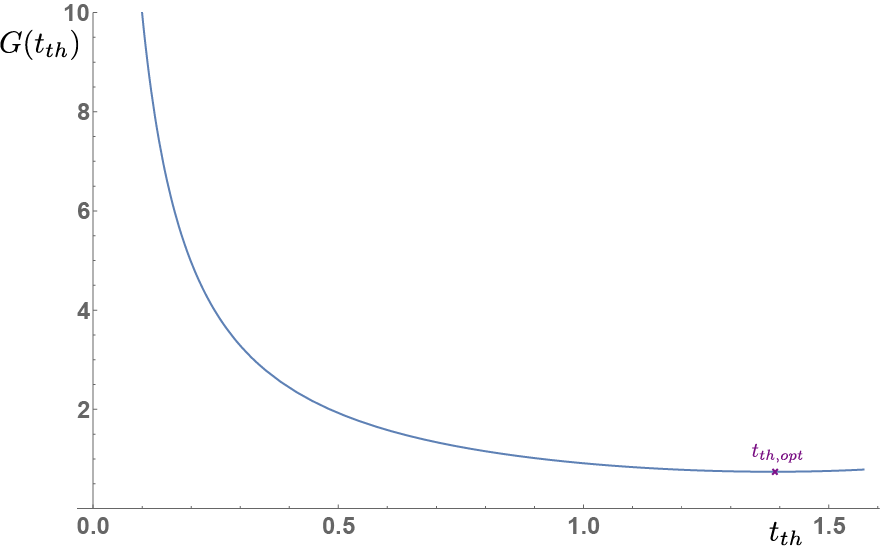}
\caption{$G(t_{\mathrm{th}})$ as a function of $t_{\mathrm{th}}$ for the relaxation model \eqref{alphatuniII}. The purple point indicates the optimal thermalization time $t_{th,opt}$.
}
\label{GI}
\end{figure}

For the collision model described in Sec. \ref{CollisionModel}, we consider \eqref{alphatuni}, i.e., 
\begin{align}
\alpha(t_\mathrm{th})= \cos^2( g t_\mathrm{th}/ \hbar),
\label{alphatuniII}
\end{align}
which arises due to the unitary evolution between S and each qubit of B for each step of the protocol. 
 Focusing on \eqref{alphatuniII}, the optimal $t_\mathrm{th}$ can be numerically found to satisfy $g t_\mathrm{th} / \hbar \approx 1.4$, for which $\alpha \approx 0.03$, very close to the value of perfect thermalization. Note that the optimal $\alpha$ is independent of $g/\hbar$. In Fig. \ref{GI}, we plot $G(t_{\mathrm{th}})$ as a function of $t_{\mathrm{th}}$ for the model \eqref{alphatuniII} with $g \hbar=1$. Clearly, we note that $G(t_{\mathrm{th}})$ leads to low dissipation when taking $t_{\mathrm{th}} \in (1,1.5)$, whereas it diverges in the limit $t_{\mathrm{th}} \rightarrow 0$. To understand this behaviour, note that for short times,
\begin{align}
\alpha(t_\mathrm{th}) \approx 1 - \mathcal{O}(t^2_\mathrm{th}),
\end{align}
and hence $W_{\rm dis} \propto t_{\rm th}^{-1}$. This behaviour is expected in circumstances where  S interacts with an independent element of B at each step, as it naturally happens in collision models. Hence we conclude that in such models it is desirable to avoid too short thermalization times to ensure low dissipation in the process.

\begin{figure}[t!]
\centering
\includegraphics[width=0.47\textwidth]{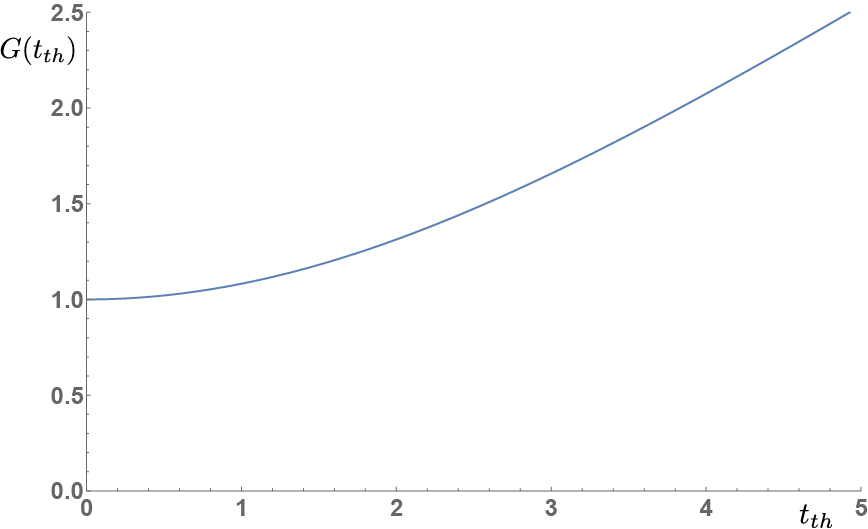}
\caption{$G(t_{\mathrm{th}})$ as a function of $t_{\mathrm{th}}$ for the relaxation model \eqref{alphatexpII}. 
}
\label{GII}
\end{figure}

We now move to a thermalization model of the form 
\begin{align}
\alpha(t_\mathrm{th}) \approx e^{-t_{\mathrm{th}}/\tau_{\mathrm{th}}},
\label{alphatexpII}
\end{align}
where $\tau_{\mathrm{th}}$ is the thermalization time-scale. This behaviour is expected for many physical systems (e.g., in open systems), at least for sufficiently large times $t_{\mathrm{th}}$. For this model of thermalization, we plot $G(t_{\mathrm{th}})$ in Fig. \ref{GII}, which shows that $G(t_{\mathrm{th}})$ increases monotonically with $t_{\mathrm{th}}$. To minimize dissipation, it is hence desirable to take small $t_{\mathrm{th}}$, as long as \eqref{alphatexpII} remains valid. Keeping that in mind, taking small but finite $t_{\mathrm{th}}$ in \eqref{Wdissopt}, we find that at leading order $W_{\rm dis} \propto \mathcal{T}^{-1}$. That is, the dissipation becomes inversely proportional to the total time in the continuous limit $t_{\mathrm{th}}\rightarrow 0$, a behaviour that is often assumed on phenomenological grounds \cite{Esposito2010C}, or obtained by means of Markovian master equations \cite{gaveau1996master,Cavina2017}.\\
 \\

Summarizing, we have observed markedly different behaviours of $W_{\rm dis}$ as a function of the thermalization time $t_{\rm th}$ in each thermalization step. While in collision-like models short $t_{\rm th}$ are not desirable, the opposite holds for an exponential relaxation as in \eqref{alphatexpII}, which is expected in situations where S is constantly coupled to B.   These considerations make a connection between isothermal processes in discrete processes, where perfect thermalization is taken for granted \cite{Anders13,AbergWork,Gallego14,Perarnau-Llobet2015a,Wilming2016,Perarnau2017}, and in continuous ones \cite{gaveau1996master,Cavina2017}.

\section{Conclusion}
\label{Conc}
In this work we have considered thermodynamic processes in which contacts between S and B correspond to imperfect thermalizations, which capture large classes of possible error models or finite-time equilibration processes. The degree of thermalization is quantified by a parameter $\alpha$, such that for $\alpha=0$ we recover the standard case of full thermalization and for $\alpha=1$ there is no interaction between S and B (and W in the collision model). We have shown  that  optimal  processes can be constructed for any $\alpha < 1$ with protocols consisting of sufficiently many steps, and characterized the corresponding tradeoff between the number of extra steps and $\alpha$. This result has been obtained in two frameworks: one based on collision models and one where the energy levels of S and interactions with B are described through quantum channels that provide an effective model for thermalization. Finally, we have discussed how optimizing the thermalisation time $t_{\rm th}$ in each thermalization step leads to notably different results for each of the considered frameworks.

\section{Acknowledgments}
We thank Guillem Perarnau, Matteo Lostaglio,  Kamil Korzekwa, and Chris Perry for interesting discussions and useful comments on the manuscript. We acknowledge contributions from the Swiss National Science Foundation via the NCCR QSIT as well as project No.\ 200020\_165843. M.P.-L. acknowledges support from the Alexander von Humboldt Foundation. All authors are grateful for support from the EU COST Action MP1209 on Thermodynamics in the Quantum Regime.
\clearpage

\bibliographystyle{unsrtnat}
\bibliography{bibliographyII} 

\begin{thebibliography}{49}
\providecommand{\natexlab}[1]{#1}
\providecommand{\url}[1]{\texttt{#1}}
\expandafter\ifx\csname urlstyle\endcsname\relax
  \providecommand{\doi}[1]{doi: #1}\else
  \providecommand{\doi}{doi: \begingroup \urlstyle{rm}\Url}\fi

\bibitem[Goold et~al.(2016)Goold, Huber, Riera, Rio, and Skrzypczyk]{Goold2016}
John Goold, Marcus Huber, Arnau Riera, L\'idia~del Rio, and Paul Skrzypczyk.
\newblock The role of quantum information in thermodynamics—a topical review.
\newblock \emph{Journal of Physics A: Mathematical and Theoretical},
  49\penalty0 (14):\penalty0 143001, Feb 2016.
\newblock ISSN 1751-8121.
\newblock \doi{10.1088/1751-8113/49/14/143001}.
\newblock URL \url{http://dx.doi.org/10.1088/1751-8113/49/14/143001}.

\bibitem[Vinjanampathy and Anders(2016)]{Vinjanampathy2016}
Sai Vinjanampathy and Janet Anders.
\newblock Quantum thermodynamics.
\newblock \emph{Contemporary Physics}, 57\penalty0 (4):\penalty0 545–579, Jul
  2016.
\newblock ISSN 1366-5812.
\newblock \doi{10.1080/00107514.2016.1201896}.
\newblock URL \url{http://dx.doi.org/10.1080/00107514.2016.1201896}.

\bibitem[Breuer and Petruccione(2006)]{Breuer2002}
H.-P Breuer and Francesco Petruccione.
\newblock \emph{The Theory of Open Quantum Systems}.
\newblock Oxford University Press, 01 2006.
\newblock \doi{10.1093/acprof:oso/9780199213900.001.0001}.

\bibitem[Ziman et~al.(2005)Ziman, {\v{S}}telmachovi{\v{c}}, and
  Bu{\v{z}}ek]{ziman2005description}
M{\"a}rio Ziman, Peter {\v{S}}telmachovi{\v{c}}, and Vladim{\'i}r Bu{\v{z}}ek.
\newblock Description of quantum dynamics of open systems based on
  collision-like models.
\newblock \emph{Open Systems {\&} Information Dynamics}, 12\penalty0
  (1):\penalty0 81--91, Mar 2005.
\newblock ISSN 1573-1324.
\newblock \doi{10.1007/s11080-005-0488-0}.
\newblock URL \url{https://doi.org/10.1007/s11080-005-0488-0}.

\bibitem[Brand\~ao et~al.(2013)Brand\~ao, Horodecki, Oppenheim, Renes, and
  Spekkens]{ResourceTheory}
Fernando G. S.~L. Brand\~ao, Micha\l{} Horodecki, Jonathan Oppenheim, Joseph~M.
  Renes, and Robert~W. Spekkens.
\newblock Resource theory of quantum states out of thermal equilibrium.
\newblock \emph{Phys. Rev. Lett.}, 111:\penalty0 250404, Dec 2013.
\newblock \doi{10.1103/PhysRevLett.111.250404}.
\newblock URL \url{https://link.aps.org/doi/10.1103/PhysRevLett.111.250404}.

\bibitem[Wilming et~al.(2016)Wilming, Gallego, and Eisert]{Wilming2016}
H.~Wilming, R.~Gallego, and J.~Eisert.
\newblock Second law of thermodynamics under control restrictions.
\newblock \emph{Phys. Rev. E}, 93:\penalty0 042126, Apr 2016.
\newblock \doi{10.1103/PhysRevE.93.042126}.
\newblock URL \url{https://link.aps.org/doi/10.1103/PhysRevE.93.042126}.

\bibitem[Lekscha et~al.(2018)Lekscha, Wilming, Eisert, and
  Gallego]{Lekscha2016}
J~Lekscha, H~Wilming, J~Eisert, and R~Gallego.
\newblock Quantum thermodynamics with local control.
\newblock \emph{Physical Review E}, 97\penalty0 (2):\penalty0 022142, 2018.
\newblock \doi{10.1103/PhysRevE.97.022142}.

\bibitem[Brown et~al.(2016)Brown, Friis, and Huber]{Brown2016}
Eric~G Brown, Nicolai Friis, and Marcus Huber.
\newblock Passivity and practical work extraction using gaussian operations.
\newblock \emph{New Journal of Physics}, 18\penalty0 (11):\penalty0 113028,
  2016.
\newblock \doi{10.1088/1367-2630/18/11/113028}.
\newblock URL \url{http://stacks.iop.org/1367-2630/18/i=11/a=113028}.

\bibitem[Friis and Huber(2018)]{Friis2017}
Nicolai Friis and Marcus Huber.
\newblock Precision and work fluctuations in gaussian battery charging.
\newblock \emph{Quantum}, 2:\penalty0 61, apr 2018.
\newblock \doi{10.22331/q-2018-04-23-61}.

\bibitem[Lostaglio et~al.(2018)Lostaglio, Alhambra, and Perry]{Lostaglio2016}
Matteo Lostaglio, {\'A}lvaro~M Alhambra, and Christopher Perry.
\newblock Elementary thermal operations.
\newblock \emph{Quantum}, 2:\penalty0 52, 2018.
\newblock \doi{10.22331/q-2018-02-08-52}.

\bibitem[Perry et~al.(2018)Perry, \ifmmode \acute{C}\else
  \'{C}\fi{}wikli\ifmmode~\acute{n}\else \'{n}\fi{}ski, Anders, Horodecki, and
  Oppenheim]{Perry2016}
Christopher Perry, Piotr \ifmmode \acute{C}\else
  \'{C}\fi{}wikli\ifmmode~\acute{n}\else \'{n}\fi{}ski, Janet Anders, Micha\l{}
  Horodecki, and Jonathan Oppenheim.
\newblock A sufficient set of experimentally implementable thermal operations
  for small systems.
\newblock \emph{Phys. Rev. X}, 8:\penalty0 041049, Dec 2018.
\newblock \doi{10.1103/PhysRevX.8.041049}.
\newblock URL \url{https://link.aps.org/doi/10.1103/PhysRevX.8.041049}.

\bibitem[Nulton et~al.(1985)Nulton, Salamon, Andresen, and Anmin]{Nulton1985}
J.~Nulton, P.~Salamon, B.~Andresen, and Qi~Anmin.
\newblock Quasistatic processes as step equilibrations.
\newblock \emph{The Journal of Chemical Physics}, 83\penalty0 (1):\penalty0
  334--338, jul 1985.
\newblock \doi{10.1063/1.449774}.
\newblock URL \url{https://doi.org/10.1063/1.449774}.

\bibitem[Anders and Giovannetti(2013)]{Anders13}
Janet Anders and Vittorio Giovannetti.
\newblock Thermodynamics of discrete quantum processes.
\newblock \emph{New Journal of Physics}, 15\penalty0 (3):\penalty0 033022, mar
  2013.
\newblock \doi{10.1088/1367-2630/15/3/033022}.
\newblock URL \url{https://doi.org/10.1088%2F1367-2630%2F15%2F3%2F033022}.

\bibitem[Skrzypczyk et~al.(2013)Skrzypczyk, Short, and Popescu]{Popescu2013b}
P.~Skrzypczyk, A.~J. Short, and S.~Popescu.
\newblock Extracting work from quantum systems.
\newblock {arXiv:1302.2811}, 2013.
\newblock URL \url{https://arxiv.org/abs/1302.2811}.

\bibitem[Reeb and Wolf(2014)]{Reeb2014}
David Reeb and Michael~M Wolf.
\newblock An improved landauer principle with finite-size corrections.
\newblock \emph{New Journal of Physics}, 16\penalty0 (10):\penalty0 103011, oct
  2014.
\newblock \doi{10.1088/1367-2630/16/10/103011}.
\newblock URL \url{https://doi.org/10.1088%2F1367-2630%2F16%2F10%2F103011}.

\bibitem[Scarani et~al.(2002)Scarani, Ziman, \ifmmode \check{S}\else
  \v{S}\fi{}telmachovi\ifmmode~\check{c}\else \v{c}\fi{}, Gisin, and
  Bu\ifmmode~\check{z}\else \v{z}\fi{}ek]{Scarani2002}
Valerio Scarani, M\'ario Ziman, Peter \ifmmode \check{S}\else
  \v{S}\fi{}telmachovi\ifmmode~\check{c}\else \v{c}\fi{}, Nicolas Gisin, and
  Vladim\'{\i}r Bu\ifmmode~\check{z}\else \v{z}\fi{}ek.
\newblock Thermalizing quantum machines: Dissipation and entanglement.
\newblock \emph{Phys. Rev. Lett.}, 88:\penalty0 097905, Feb 2002.
\newblock \doi{10.1103/PhysRevLett.88.097905}.
\newblock URL \url{https://link.aps.org/doi/10.1103/PhysRevLett.88.097905}.

\bibitem[Filipowicz et~al.(1986)Filipowicz, Javanainen, and
  Meystre]{Filipowicz1986}
P.~Filipowicz, J.~Javanainen, and P.~Meystre.
\newblock Theory of a microscopic maser.
\newblock \emph{Phys. Rev. A}, 34:\penalty0 3077--3087, Oct 1986.
\newblock \doi{10.1103/PhysRevA.34.3077}.
\newblock URL \url{https://link.aps.org/doi/10.1103/PhysRevA.34.3077}.

\bibitem[Caves and Milburn(1987)]{Caves1987}
Carlton~M. Caves and G.~J. Milburn.
\newblock Quantum-mechanical model for continuous position measurements.
\newblock \emph{Phys. Rev. A}, 36:\penalty0 5543--5555, Dec 1987.
\newblock \doi{10.1103/PhysRevA.36.5543}.
\newblock URL \url{https://link.aps.org/doi/10.1103/PhysRevA.36.5543}.

\bibitem[\AA{}berg(2014)]{Aberg2013}
Johan \AA{}berg.
\newblock Catalytic coherence.
\newblock \emph{Phys. Rev. Lett.}, 113:\penalty0 150402, Oct 2014.
\newblock \doi{10.1103/PhysRevLett.113.150402}.
\newblock URL \url{https://link.aps.org/doi/10.1103/PhysRevLett.113.150402}.

\bibitem[\AA{}berg(2013)]{AbergWork}
J.~\AA{}berg.
\newblock Truly work-like work extraction via a single-shot analysis.
\newblock \emph{Nat. Commun.}, 4\penalty0 (1925):\penalty0 1925, 2013.
\newblock \doi{10.1038/ncomms2712}.
\newblock URL \url{https://www.nature.com/articles/ncomms2712}.

\bibitem[Ziman et~al.(2002)Ziman, \ifmmode \check{S}\else
  \v{S}\fi{}telmachovi\ifmmode~\check{c}\else \v{c}\fi{},
  Bu\ifmmode~\check{z}\else \v{z}\fi{}ek, Hillery, Scarani, and
  Gisin]{Ziman2002}
M.~Ziman, P.~\ifmmode \check{S}\else
  \v{S}\fi{}telmachovi\ifmmode~\check{c}\else \v{c}\fi{},
  V.~Bu\ifmmode~\check{z}\else \v{z}\fi{}ek, M.~Hillery, V.~Scarani, and
  N.~Gisin.
\newblock Diluting quantum information: An analysis of information transfer in
  system-reservoir interactions.
\newblock \emph{Phys. Rev. A}, 65:\penalty0 042105, Mar 2002.
\newblock \doi{10.1103/PhysRevA.65.042105}.
\newblock URL \url{https://link.aps.org/doi/10.1103/PhysRevA.65.042105}.

\bibitem[Gennaro et~al.(2009)Gennaro, Benenti, and Palma]{Gennaro2015}
Giuseppe Gennaro, Giuliano Benenti, and G.~Massimo Palma.
\newblock Relaxation due to random collisions with a many-qudit environment.
\newblock \emph{Phys. Rev. A}, 79:\penalty0 022105, Feb 2009.
\newblock \doi{10.1103/PhysRevA.79.022105}.
\newblock URL \url{https://link.aps.org/doi/10.1103/PhysRevA.79.022105}.

\bibitem[Cusumano et~al.(2018)Cusumano, Cavina, Keck, De~Pasquale, and
  Giovannetti]{cusumano2018entropy}
Stefano Cusumano, Vasco Cavina, Maximilian Keck, Antonella De~Pasquale, and
  Vittorio Giovannetti.
\newblock Entropy production and asymptotic factorization via thermalization: A
  collisional model approach.
\newblock \emph{Phys. Rev. A}, 98:\penalty0 032119, Sep 2018.
\newblock \doi{10.1103/PhysRevA.98.032119}.
\newblock URL \url{https://link.aps.org/doi/10.1103/PhysRevA.98.032119}.

\bibitem[Man et~al.(2018)Man, Xia, and Lo~Franco]{Man2018}
Zhong-Xiao Man, Yun-Jie Xia, and Rosario Lo~Franco.
\newblock Temperature effects on quantum non-markovianity via collision models.
\newblock \emph{Phys. Rev. A}, 97:\penalty0 062104, Jun 2018.
\newblock \doi{10.1103/PhysRevA.97.062104}.
\newblock URL \url{https://link.aps.org/doi/10.1103/PhysRevA.97.062104}.

\bibitem[Di{\'o}si et~al.(2006)Di{\'o}si, Feldmann, and
  Kosloff]{diosi2006exact}
Lajos Di{\'o}si, Tova Feldmann, and Ronnie Kosloff.
\newblock On the exact identity between thermodynamic and informatic entropies
  in a unitary model of friction.
\newblock \emph{International Journal of Quantum Information}, 4\penalty0
  (01):\penalty0 99--104, 2006.
\newblock \doi{10.1142/S0219749906001645}.
\newblock URL
  \url{https://www.worldscientific.com/doi/abs/10.1142/S0219749906001645}.

\bibitem[Uzdin and Kosloff(2014)]{Raam2014}
Raam Uzdin and Ronnie Kosloff.
\newblock The multilevel four-stroke swap engine and its environment.
\newblock \emph{New Journal of Physics}, 16\penalty0 (9):\penalty0 095003,
  2014.
\newblock \doi{10.1088/1367-2630/16/9/095003}.
\newblock URL \url{http://stacks.iop.org/1367-2630/16/i=9/a=095003}.

\bibitem[Barra(2015)]{barra2015thermodynamic}
Felipe Barra.
\newblock The thermodynamic cost of driving quantum systems by their
  boundaries.
\newblock \emph{Scientific reports}, 5\penalty0 (14873):\penalty0 14873, 2015.
\newblock \doi{10.1038/srep14873}.
\newblock URL \url{https://www.nature.com/articles/srep14873}.

\bibitem[Lorenzo et~al.(2015)Lorenzo, McCloskey, Ciccarello, Paternostro, and
  Palma]{Lorenzo2015}
S.~Lorenzo, R.~McCloskey, F.~Ciccarello, M.~Paternostro, and G.~M. Palma.
\newblock Landauer's principle in multipartite open quantum system dynamics.
\newblock \emph{Phys. Rev. Lett.}, 115:\penalty0 120403, Sep 2015.
\newblock \doi{10.1103/PhysRevLett.115.120403}.
\newblock URL \url{https://link.aps.org/doi/10.1103/PhysRevLett.115.120403}.

\bibitem[Pezzutto et~al.(2016)Pezzutto, Paternostro, and
  Omar]{pezzutto2016implications}
Marco Pezzutto, Mauro Paternostro, and Yasser Omar.
\newblock Implications of non-markovian quantum dynamics for the landauer
  bound.
\newblock \emph{New Journal of Physics}, 18\penalty0 (12):\penalty0 123018, dec
  2016.
\newblock \doi{10.1088/1367-2630/18/12/123018}.
\newblock URL \url{https://doi.org/10.1088%2F1367-2630%2F18%2F12%2F123018}.

\bibitem[Strasberg et~al.(2017)Strasberg, Schaller, Brandes, and
  Esposito]{Strasberg2017}
Philipp Strasberg, Gernot Schaller, Tobias Brandes, and Massimiliano Esposito.
\newblock Quantum and information thermodynamics: A unifying framework based on
  repeated interactions.
\newblock \emph{Phys. Rev. X}, 7:\penalty0 021003, Apr 2017.
\newblock \doi{10.1103/PhysRevX.7.021003}.
\newblock URL \url{https://link.aps.org/doi/10.1103/PhysRevX.7.021003}.

\bibitem[Skrzypczyk et~al.(2014)Skrzypczyk, Short, and Popescu]{Popescu2013}
P.~Skrzypczyk, A.~J. Short, and S.~Popescu.
\newblock Work extraction and thermodynamics for individual quantum systems.
\newblock \emph{Nature Comm.}, 5\penalty0 (4185):\penalty0 4185, 2014.
\newblock \doi{10.1038/ncomms5185}.
\newblock URL \url{https://www.nature.com/articles/ncomms5185}.

\bibitem[Horodecki and Oppenheim(2013)]{Nanomachines}
M.~Horodecki and J.~Oppenheim.
\newblock Fundamental limitations for quantum and nanoscale thermodynamics.
\newblock \emph{Nature Comm.}, 4\penalty0 (2059):\penalty0 033022, 2013.
\newblock \doi{10.1038/ncomms3059}.
\newblock URL \url{https://www.nature.com/articles/ncomms3059}.

\bibitem[Gour et~al.(2015)Gour, Muller, Narasimhachar, Spekkens, and
  Halpern]{Gour13}
Gilad Gour, Markus~P. Muller, Varun Narasimhachar, Robert~W. Spekkens, and
  Nicole~Yunger Halpern.
\newblock The resource theory of informational nonequilibrium in
  thermodynamics.
\newblock \emph{Physics Reports}, 583:\penalty0 1 -- 58, 2015.
\newblock ISSN 0370-1573.
\newblock \doi{10.1016/j.physrep.2015.04.003}.
\newblock URL
  \url{http://www.sciencedirect.com/science/article/pii/S037015731500229X}.

\bibitem[M\"uller(2018)]{mueller2017correlating}
Markus~P. M\"uller.
\newblock Correlating thermal machines and the second law at the nanoscale.
\newblock \emph{Phys. Rev. X}, 8:\penalty0 041051, Dec 2018.
\newblock \doi{10.1103/PhysRevX.8.041051}.
\newblock URL \url{https://link.aps.org/doi/10.1103/PhysRevX.8.041051}.

\bibitem[Korzekwa et~al.(2016)Korzekwa, Lostaglio, Oppenheim, and
  Jennings]{Korzekwa2016}
Kamil Korzekwa, Matteo Lostaglio, Jonathan Oppenheim, and David Jennings.
\newblock The extraction of work from quantum coherence.
\newblock \emph{New Journal of Physics}, 18\penalty0 (2):\penalty0 023045, feb
  2016.
\newblock \doi{10.1088/1367-2630/18/2/023045}.
\newblock URL \url{https://doi.org/10.1088%2F1367-2630%2F18%2F2%2F023045}.

\bibitem[Alicki et~al.(2004)Alicki, Horodecki, Horodecki, and
  Horodecki]{Alicki2004}
Robert Alicki, Michał Horodecki, Paweł Horodecki, and Ryszard Horodecki.
\newblock Thermodynamics of quantum information systems — hamiltonian
  description.
\newblock \emph{Open Systems Information Dynamics}, 11\penalty0 (03):\penalty0
  205--217, 2004.
\newblock \doi{10.1023/B:OPSY.0000047566.72717.71}.
\newblock URL
  \url{http://www.worldscientific.com/doi/abs/10.1023/B3AOPSY.0000047566.72717.71}.

\bibitem[Esposito and den Broeck(2011)]{Esposito2011}
M.~Esposito and C.~Van den Broeck.
\newblock Second law and landauer principle far from equilibrium.
\newblock \emph{{EPL} (Europhysics Letters)}, 95\penalty0 (4):\penalty0 40004,
  aug 2011.
\newblock \doi{10.1209/0295-5075/95/40004}.
\newblock URL \url{https://doi.org/10.1209%2F0295-5075%2F95%2F40004}.

\bibitem[Talkner et~al.(2007)Talkner, Lutz, and H\"anggi]{Talkner07}
Peter Talkner, Eric Lutz, and Peter H\"anggi.
\newblock Fluctuation theorems: Work is not an observable.
\newblock \emph{Phys. Rev. E}, 75:\penalty0 050102, May 2007.
\newblock \doi{10.1103/PhysRevE.75.050102}.
\newblock URL \url{https://link.aps.org/doi/10.1103/PhysRevE.75.050102}.

\bibitem[Perarnau-Llobet et~al.(2018)Perarnau-Llobet, Wilming, Riera, Gallego,
  and Eisert]{Perarnau2017}
M.~Perarnau-Llobet, H.~Wilming, A.~Riera, R.~Gallego, and J.~Eisert.
\newblock Strong coupling corrections in quantum thermodynamics.
\newblock \emph{Physical Review Letters}, 120\penalty0 (12), mar 2018.
\newblock \doi{10.1103/physrevlett.120.120602}.

\bibitem[Scandi and Perarnau-Llobet(2018)]{scandi2018thermodynamic}
Matteo Scandi and Mart{\'\i} Perarnau-Llobet.
\newblock Thermodynamic length in open quantum systems.
\newblock \emph{arXiv preprint arXiv:1810.05583}, 2018.
\newblock URL \url{https://arxiv.org/abs/1810.05583}.

\bibitem[Esposito et~al.(2010{\natexlab{a}})Esposito, Lindenberg, and den
  Broeck]{Esposito2010}
Massimiliano Esposito, Katja Lindenberg, and Christian~Van den Broeck.
\newblock Entropy production as correlation between system and reservoir.
\newblock \emph{New Journal of Physics}, 12\penalty0 (1):\penalty0 013013, jan
  2010{\natexlab{a}}.
\newblock \doi{10.1088/1367-2630/12/1/013013}.
\newblock URL \url{https://doi.org/10.1088%2F1367-2630%2F12%2F1%2F013013}.

\bibitem[Gallego et~al.(2014)Gallego, Riera, and Eisert]{Gallego14}
R~Gallego, A~Riera, and J~Eisert.
\newblock Thermal machines beyond the weak coupling regime.
\newblock \emph{New Journal of Physics}, 16\penalty0 (12):\penalty0 125009, dec
  2014.
\newblock \doi{10.1088/1367-2630/16/12/125009}.
\newblock URL \url{https://doi.org/10.1088%2F1367-2630%2F16%2F12%2F125009}.

\bibitem[Jacobs(2009)]{Jacobs2009}
Kurt Jacobs.
\newblock Second law of thermodynamics and quantum feedback control: Maxwell's
  demon with weak measurements.
\newblock \emph{Phys. Rev. A}, 80:\penalty0 012322, Jul 2009.
\newblock \doi{10.1103/PhysRevA.80.012322}.
\newblock URL \url{https://link.aps.org/doi/10.1103/PhysRevA.80.012322}.

\bibitem[Perarnau-Llobet et~al.(2017)Perarnau-Llobet, B\"aumer, Hovhannisyan,
  Huber, and Acin]{Perarnau-Llobet2016}
Mart\'{\i} Perarnau-Llobet, Elisa B\"aumer, Karen~V. Hovhannisyan, Marcus
  Huber, and Antonio Acin.
\newblock No-go theorem for the characterization of work fluctuations in
  coherent quantum systems.
\newblock \emph{Phys. Rev. Lett.}, 118:\penalty0 070601, Feb 2017.
\newblock \doi{10.1103/PhysRevLett.118.070601}.
\newblock URL \url{https://link.aps.org/doi/10.1103/PhysRevLett.118.070601}.

\bibitem[Esposito et~al.(2010{\natexlab{b}})Esposito, Kawai, Lindenberg, and
  Van~den Broeck]{Esposito2010C}
Massimiliano Esposito, Ryoichi Kawai, Katja Lindenberg, and Christian Van~den
  Broeck.
\newblock Efficiency at maximum power of low-dissipation carnot engines.
\newblock \emph{Phys. Rev. Lett.}, 105:\penalty0 150603, Oct
  2010{\natexlab{b}}.
\newblock \doi{10.1103/PhysRevLett.105.150603}.
\newblock URL \url{https://link.aps.org/doi/10.1103/PhysRevLett.105.150603}.

\bibitem[Gaveau and Schulman(1996)]{gaveau1996master}
Bernard Gaveau and LS~Schulman.
\newblock Master equation based formulation of nonequilibrium statistical
  mechanics.
\newblock \emph{Journal of Mathematical Physics}, 37\penalty0 (8):\penalty0
  3897--3932, 1996.
\newblock \doi{10.1063/1.531608}.
\newblock URL \url{https://aip.scitation.org/doi/10.1063/1.531608}.

\bibitem[Cavina et~al.(2017)Cavina, Mari, and Giovannetti]{Cavina2017}
Vasco Cavina, Andrea Mari, and Vittorio Giovannetti.
\newblock Slow dynamics and thermodynamics of open quantum systems.
\newblock \emph{Phys. Rev. Lett.}, 119:\penalty0 050601, Aug 2017.
\newblock \doi{10.1103/PhysRevLett.119.050601}.
\newblock URL \url{https://link.aps.org/doi/10.1103/PhysRevLett.119.050601}.

\bibitem[Perarnau-Llobet et~al.(2016)Perarnau-Llobet, Riera, Gallego, Wilming,
  and Eisert]{Perarnau-Llobet2015a}
Mart{\'{\i}} Perarnau-Llobet, Arnau Riera, Rodrigo Gallego, Henrik Wilming, and
  Jens Eisert.
\newblock Work and entropy production in generalised gibbs ensembles.
\newblock \emph{New Journal of Physics}, 18\penalty0 (12):\penalty0 123035, dec
  2016.
\newblock \doi{10.1088/1367-2630/aa4fa6}.
\newblock URL \url{https://doi.org/10.1088%2F1367-2630%2Faa4fa6}.

\bibitem[Jarzynski(1997)]{Jarzynski97}
C.~Jarzynski.
\newblock Nonequilibrium equality for free energy differences.
\newblock \emph{Phys. Rev. Lett.}, 78:\penalty0 2690--2693, Apr 1997.
\newblock \doi{10.1103/PhysRevLett.78.2690}.
\newblock URL \url{https://link.aps.org/doi/10.1103/PhysRevLett.78.2690}.

\end{thebibliography}

\onecolumn
\newpage

\appendix
\section*{Appendices}

\section{Models of Noise}\label{appA}

In the $k^{\rm th}$ step of the protocol, the  joint state of S, W and the $k^{\rm th}$ bath qubit can be written as,  
\begin{align}
\begin{split}
\rho^{(k)}_0 &= \int dx \hspace{1mm} p_{\rm W}^{(k)}(x) \left( \phi_X^{(\rm deg)} +\phi_X^{(\text{non-deg})}\right), \\ {\rm with} \quad \phi_X^{(\rm deg)}&=   r_k \ket{0,1,x} \bra{0,1,x}+ s_k \ket{1,0,x+\omega_k} \bra{1,0,x+\omega_k},
\\
\phi_X^{(\text{non-deg})} &=\frac{r_k s_k}{q_k p_{k-1}} \ket{0,0,x} \bra{0,0,x}+ \frac{r_k s_k}{(1-q_k) (1-p_{k-1})} \ket{1,1,x+\omega_k} \bra{1,1,x+\omega_k},
\end{split}
\end{align}
where $r_k:=q_k(1-p_{k-1})$, $s_k:=(1-q_k)p_{k-1}$, and note that $ r_k>s_k$. $p_{\rm W}^{(k)}(x)$ is a probability distribution that describes the population of the different work storage levels $\ket{x}\bra{x}_W$ after the $k^{th}$ step. The non-normalized  states $\phi_X^{(\rm deg)}$ represent  the subspaces of relevant degenerate states where we act upon for a given $x$, and $\phi_X^{(\text{non-deg})}$ are the rest of states.
 Now we consider thermal operations acting only on each $\phi_X^{(\rm deg)}$, 
\begin{align}
\psi_X^{(\rm deg)}=\Tr_A \left(V_k \phi_X^{(\rm deg)} \otimes \tau_A V_k^{\dagger} \right)
\label{arbitraryV}
\end{align}
where $\tau_A=e^{-\beta H_A}/\Tr(e^{-\beta H_A})$ is an arbitrary thermal state.
From the second law of thermodynamics, we know that the free energy of $\phi_X^{(\rm deg)} $ can only decrease which here  implies that the entropy of each $\phi_X^{(\rm deg)}$ must increase --recall that $\phi_X^{\rm deg}$ has a trivial Hamiltonian as it lives in an energy degenerate subspace.
Hence we can write without loss of generality
\begin{align}
\psi_X^{(\rm deg)}= \begin{pmatrix}
(1-\alpha_k(x))s_k+\alpha_k(x) r_k& z_k(x) \\
z^*_k(x) & (1-\alpha_k(x))r_k+\alpha_k(x) s_k
\end{pmatrix},
\end{align}
where $z_k(x)$ and $\alpha_k(x)$ are in general functions of the specific $\tau_A$ and $V_k$,  and $\alpha_k(x)\in[0,1]$. On the other hand, the total state after the thermal operation reads
\begin{align}
\rho^{(k)} &= \int dx \hspace{1mm} p_{\rm W}^{(k)}(x) \left( \psi_X^{(\rm deg)} +\phi_X^{(\text{non-deg})}\right).
\end{align}
For future convenience, 
let us also define 
\begin{align}
\alpha_k:= \int dx p_{\rm W}^{(k)}(x) \alpha_k(x).
\end{align} 
Note that, for a specific choice of $V$, where $z_k(x)=\sqrt{\alpha_k(x)(1-\alpha_k(x))}(r_k+s_k)$, we would get exactly the same result as after the application of \eqref{tUi}. Another "extreme" case would be for $z_k(x)=0$, where the resulting state on SBW would be diagonal, corresponding to the resulting state of the map 
\begin{align}
\rho \longrightarrow (1-\alpha_k)U_k \rho U_k^{\dagger}+\alpha_k \rho .
\label{tUi2}
\end{align}
That is: with a given probability $ (1-\alpha_k)$, we perform the desired operation, and with an error rate $\alpha_k$ we do nothing.\\

At the level of S, i.e., after tracing out also the bath qubit B and W, the coherence term $z_k(x)$ becomes irrelevant, and analogous to \eqref{modelnoiseIII} the final state of S is given by
\begin{align}
\rho_{\rm S}^{(k)}&=\Tr_{\rm BW} \left[\rho^{(k)}\right] \nonumber\\
&=\int dx \hspace{1mm} p_{\rm W}^{(k)}(x) \bigg[\big( (1-\alpha_k(x))s_k+\alpha_k(x) r_k  + (1-p_{k-1})(1-q_k) \big) \ket{0} \bra{0}_{\rm S} \nonumber\\
&\quad +\big( (1-\alpha_k(x))r_k+\alpha_k(x) s_k + p_{k-1}q_k \big) \ket{1}\bra{1}_{\rm S} \bigg] \nonumber\\
&=\big( (1-\alpha_k)(1-q_k)+ \alpha_k(1-p_{k-1}) \big) \ket{0} \bra{0}_{\rm S}+\big((1-\alpha_k)q_k+\alpha_k p_{k-1} \big) \ket{1}\bra{1}_{\rm S} \nonumber\\
&=(1-\alpha_k)\tau^{(k)}_{\rm B} + \alpha_k \rho_{\rm S}^{(k-1)}. \label{finalstate}
\end{align} 
Hence, for any unitary $V_k$ this more general model gives the same description of the resulting system state as the specific model we considered before \eqref{tUi}, where $V_k$ only determines the value of $\alpha_k$. In any non-trivial unitary operation $V_k$, i.e., for any $\alpha<1$, S and B get mixed, meaning that it brings the state of S closer to the thermal state of B. 
Analogously, when tracing out the system S and work battery W, we get the same evolution for the $k^{\rm th}$ bath qubit, 
\begin{align}
\begin{split}
\rho^{(k)}_{\rm S}&= \alpha_k \rho^{(k-1)}_{\rm S} + (1-\alpha_k) \tau_{\rm B}^{(k)}
\\
\rho_{\rm B}^{(k)} &= \alpha_k \tau_{\rm B}^{(k)} + (1-\alpha_k) \rho^{(k-1)}_{\rm S}.
\end{split}
\end{align}
Here, $\rho_{\rm S}^{(k)}$ and $\rho_{\rm B}^{(k)}$ represent the final state of S and B after applying the $k^{\rm th}$ unitary operation.

Finally, we stress that the unitary $V$ does not encompass the most general energy preserving unitary on SBWA. In particular, we note that in our previous considerations S interacts successively with each bath qubit and always on all three systems S, B and W together. This sequence is kept as we are interested here in quasistatic isothermal processes.

\section{Discussion of Noise Types that the Protocol Cannot Tolerate}\label{appB}
Here we can briefly discuss other imperfect operations which will lead to non-ideality of the process. 
The error models discussed previously are not the most general energy conserving unitaries. Indeed, since we assumed that W has a continuous Hamiltonian, there are many other degenerate subspaces. Therefore, there are in principle other energy-preserving operations, namely
\begin{itemize}
\item First of all, note that we assume that in each step the correct bath qubit is chosen, i.e., the system interacts successively with each of the $N$ bath qubits, such that we a have quasistatic isothermal process. Changing considerably the order would lead to  energy-preserving dissipative processes.
\item There can  be energy-preserving interactions  between S and W only, which would on average excite the system and therefore decrease the extracted work. The same would happen for random interactions between B and W, where in addition the correlations between S and W would be destroyed leading to uncontrolled fluctuations.
\item  When considering all three systems S, B and W together, there is not only the degeneracy between $\ket{0}_{\rm S}\ket{1}_{\rm B}\ket{x}_{\rm W} \leftrightarrow \ket{1}_{\rm S}\ket{0}_{\rm B}\ket{x+E_{k}-\epsilon_{\rm S}}_{\rm W}$, but there would also be a degeneracy between $\ket{0}_{\rm S}\ket{0}_{\rm B}\ket{x}_{\rm W} \leftrightarrow \ket{1}_{\rm S}\ket{1}_{\rm B}\ket{x-E_{k}-\epsilon_{\rm S}}_{\rm W}$ due to the continuous Hamiltonian $H_{\rm W}$. Since starting in $\ket{0}_{\rm S}\ket{0}_{\rm B}\ket{x}_{\rm W}$ is much more likely than starting in $\ket{1}_{\rm S}\ket{1}_{\rm B}\ket{x-E_{k}-\epsilon_{\rm S}}$, we would also loose work on average during this transition. 
\end{itemize}
We can see that errors on all other kind of degeneracies prevent us from extracting maximal work. The only degeneracy on which errors can be compensated for is essentially the one on which the ideal protocol is interacting. 
While these considerations provide a qualitative analysis on the kind of errors that can/cannot be tolerated, we leave as an interesting task for the future to characterize a scenario where both kind of errors are present simultaneously.

\section{Fluctuations}
\label{appFluct}
In order to calculate the fluctuations, we need to list each possible path with its corresponding probability and energy. We use vectors $\pvec{k}{0}$ and $\pvec{k}{1}$ in which the probabilities of all paths that end in state 0, or 1, respectively, are listed as well as vectors $\evec{k}{0}$ and $\evec{k}{1}$ that list the corresponding energies. With each step, the number of possible paths doubles, such that after $k$ steps, we will have $2^{k}$-dimensional vectors.\\
As we have shown before, any of our considered noise models can be seen as "apply a swap" with probability $(1-\alpha)$ and ``apply identity" with probability $\alpha$. Using \eqref{modelnoise}, we can find a recursive way to describe the vectors:
\begin{align}
\begin{split}
\pvec{k}{0}&=\begin{pmatrix}
\pvec{k-1}{0} \left( (1-q_k)+\alpha q_k \right) \\
\pvec{k-1}{1} (1-\alpha)(1-q_k)
\end{pmatrix}, \qquad
\evec{k}{0}=\begin{pmatrix}
\evec{k-1}{0}  \\
\evec{k-1}{1} -E_k \mathbb{I} 
\end{pmatrix} , \\
\pvec{k}{1}&=\begin{pmatrix}
\pvec{k-1}{1} \left( q_k+\alpha (1-q_k) \right) \\
\pvec{k-1}{0} (1-\alpha)q_k
\end{pmatrix},\qquad
\evec{k}{1}=\begin{pmatrix}
\evec{k-1}{1}  \\
\evec{k-1}{0} +E_k \mathbb{I} 
\end{pmatrix},\\
\text{with} \qquad \pvec{0}{0}&= \begin{pmatrix} 1\end{pmatrix}, \quad \pvec{0}{1}= (), \quad \evec{0}{0}= \begin{pmatrix} 0\end{pmatrix}, \quad \evec{0}{1}= (),
\end{split}
\end{align}
where $q_k$ corresponds to the excitation probability of the $k^{th}$ bath qubit and $E_k=\ln \left( \frac{1-q_k}{q_k}\right)$ to the corresponding energy gap.\\
We can calculate the average work after the $m^{th}$ step,
\begin{align}
\ave{m}&=\pvec{m}{0} \cdot \evec{m}{0}+\pvec{m}{1} \cdot \evec{m}{1}\nonumber\\
&= \pvec{m-1}{0} \cdot \evec{m-1}{0}+\pvec{m-1}{1} \cdot \evec{m-1}{1} +(1-\alpha)E_m\left(q_m\norm{\pvec{m-1}{0}}-(1-q_m)\norm{\pvec{m-1}{1}}\right)\nonumber\\
&=\ave{m-1}+(1-\alpha)E_m\left(q_m-\norm{\pvec{m-1}{1}}\right)\nonumber\\
&=(1-\alpha)\sum_{i=1}^mE_i\left(q_i-\norm{\pvec{i-1}{1}}\right),
\end{align}
as well as the square of it, since this is needed for calculating the variance,
\begin{align}
\ave{m}^2&=\ave{m-1}^2+(1-\alpha)^2E_m^2\left(q_m-\norm{\pvec{m-1}{1}}\right)^2+2(1-\alpha)\ave{m-1}E_m\left(q_m-\norm{\pvec{m-1}{1}}\right) \nonumber\\
&=\sum_{i=1}^m (1-\alpha)^2E_i^2\left(q_i-\norm{\pvec{i-1}{1}}\right)^2+2(1-\alpha)\ave{i-1}E_i\left(q_i-\norm{\pvec{i-1}{1}}\right).
\end{align}
Next, we calculate the sum of all elements of $\pvec{k}{1}$, which corresponds to the excitation probability $p_k$ of the system qubit after the $k^{\rm th}$ step:
\begin{align}
\norm{\pvec{k}{1}}&=\norm{\pvec{k-1}{1}}\left(q_k+\alpha(1-q_k)\right)+\norm{\pvec{k-1}{0}}(1-\alpha)q_k\nonumber\\
&=(1-\alpha)q_k\left(\norm{\pvec{k-1}{1}}+\norm{\pvec{k-1}{0}}\right)+\alpha \norm{\pvec{k-1}{1}}\nonumber\\
&=(1-\alpha)q_k+\alpha \norm{\pvec{k-1}{1}}
\end{align}
Using this recursive form, we can determine the explicit form:
\begin{align}
p_k=\norm{\pvec{k}{1}}=(1-\alpha)\sum_{i=1}^{k} \alpha^{k-i}q_i+\alpha^k q_0,
\end{align}
which gives us the same result as \eqref{explicit}.
In addition, we need the average value of the squared work after $m$ steps, where in the following $\vec{\bf{W}}_{m}^{2\ [1]}$ denotes the entry-wise square product of each vector element
\begin{align}
\langle W_{m}^2\rangle &= \pvec{m}{0} \cdot \vec{\bf{W}}_{m}^{2\ [0]}+ \pvec{m}{1} \cdot \vec{\bf{W}}_{m}^{2\ [1]}\nonumber \\
&=\pvec{m-1}{0} \cdot \vec{\bf{W}}_{m-1}^{2\ [0]}+ \pvec{m-1}{1} \cdot \vec{\bf{W}}_{m-1}^{2\ [1]} \nonumber \\
&\quad + \left(E_m^2\mathbb{I} -2E_m \evec{m-1}{1}\right)(1-\alpha)(1-q_m)\pvec{m-1}{1}+\left(E_m^2\mathbb{I}+2E_m\evec{m-1}{0} \right) (1-\alpha)q_m\pvec{m-1}{0} \nonumber \\
&=\langle W_{m-1}^2\rangle \!+\!2 E_m(1\!-\!\alpha)\! \left( q_m\ave{m-1} \!-\! \pvec{m-1}{1} \evec{m-1}{1}\right)\!+\!E_m^2(1-\alpha)\left(q_m\!+\!\norm{\pvec{m-1}{1}}\!-\!2q_m\norm{\pvec{m-1}{1}}\right)\nonumber \\
&=\sum_{i=1}^m 2 E_i(1-\alpha) \left( q_i\ave{i-1} -\pvec{i-1}{1} \evec{i-1}{1}\right)+E_i^2(1-\alpha)\left(q_i+\norm{\pvec{i-1}{1}}-2q_i\norm{\pvec{i-1}{1}}\right).
\end{align}
Now we can calculate the variance after the total $N$ steps as
\begin{align}
\mathrm{Var}(W_N)&=\langle W_{N}^2\rangle - \ave{N}^2\nonumber\\
\begin{split}\label{var1} &=\sum_{m=1}^N E_m^2(1-\alpha)\left(q_m(1-q_m)+\norm{\pvec{m-1}{1}}(1-\norm{\pvec{m-1}{1}})+ \alpha (q_m-\pvec{m-1}{1})^2\right)  \\
&\quad \quad +2E_m(1-\alpha)\left(\norm{\pvec{m-1}{1}} \ave{m-1}-\pvec{m-1}{1}\evec{m-1}{1}\right)\end{split}
\end{align}
Writing the last term explicitly, which determines the average work after $m$ steps when ending in the excited state $\ket{1}_{\rm S}$, we get
\begin{align}
\pvec{m}{1} \evec{m}{1} &= \pvec{m-1}{1} \evec{m-1}{1} \left(q_m + \alpha (1-q_m)\right) + \pvec{m-1}{0} \left(\evec{m-1}{0}+E_m\mathbb{I}\right) (1-\alpha)q_m\nonumber\\
&=q_m(1-\alpha)\left(\ave{m-1}+(1-\norm{\pvec{m-1}{1}})E_m\right)+\alpha \pvec{m-1}{1} \evec{m-1}{1}\nonumber\\
&=(1-\alpha) \sum_{k=1}^m \alpha^{m-k}q_k \left(\ave{k-1}+(1-\norm{\pvec{k-1}{1}})E_k\right)\label{pmwm}.
\end{align}
To simplify these expressions and show that the variance indeed goes to zero for $N \to \infty$, we need to choose a probability distribution $q_k$ for the excitation probabilities of the bath qubits. For this proof we will use $q_k=\frac{k}{2N}$ and the corresponding energy levels are then given by $E_k=k_BT\ln\left( \frac{1-q_k}{q_k} \right) =k_BT\ln\left( \frac{2N-k}{k} \right) $, but looking at simulations, it should also work for all the other distributions allowed in this paper (i.e., those mentioned in Appendix~\ref{appGen} for the case of two-level systems). \\
First, we will simplify some expressions that will help us later on:
\begin{align}
\label{pvec}&\bullet \quad \norm{\pvec{k}{1}}=(1-\alpha)\sum_{i=1}^k \alpha^{k-i}\frac{i}{2N}=\frac{k}{2N} - \frac{\alpha}{1-\alpha} \frac{1-\alpha^k}{2N}\\
\label{diff1}&\bullet \quad q_k-\norm{\pvec{k-1}{1}}=\frac{1}{2N}\frac{1-\alpha^k}{1-\alpha} \\
\label{averagework}&\bullet \quad  \ave{N}=(1-\alpha)\sum_{i=1}^N E_k\left(q_k-\norm{\pvec{k-1}{1}}\right)=\frac{1}{2N}\sum_{k=1}^N E_k (1-\alpha^k).\\
\label{finalsum}&\bullet \quad \frac{1}{N^2} \sum_{m=1}^N E_m^2 = \frac{1}{N^2} \sum_{m=1}^N \mathcal{O}\left(\ln^2{(N)}\right)=\mathcal{O}\left( \frac{\ln^2 N}{N} \right)\\
&\bullet \quad E_m\!-\!E_{m+1}=k_BT\ln\!\left( \frac{(2N\!-\!m)(m\!+\!1)}{m(2N\!-\!m\!-\!1)} \right)\!=\!k_BT\ln\!\left(\!1\!+\!\frac{1}{2N\!-\!m\!-\!1}\right) \!+\! k_BT\ln\!\left( 1\!+\!\frac{1}{m}\right)= \ord{\frac{1}{m}}\\
&\bullet  \quad E_m^2\!-\!E_{m+1}^2\!=\!k_BT\!\ln^2\!\left( \!\frac{2N\!-\!m}{m}\right) \!-\! k_B T\! \bigg(\! \ln\!\left(\! \frac{2N\!-\!m}{m}\!\right)\!+\! \underbrace{\ln\!\left(\!\frac{2N\!-\!m\!-\!1}{2N\!-\!m}\right)\!}_{=\ord{\frac{1}{N}}}\!+\! \underbrace{\ln\left( \frac{m}{m+1}\right)\!}_{=\ord{\frac{1}{m}}}\!\bigg)^2\!= \!\ord{\!\frac{\ln N}{m}}\!,
\end{align}
and estimate the sum:
\begin{align}
\sum_{i=k}^m E_i&=k_BT\ln\left(\prod_{i=k}^m \frac{2N-i}{i}\right) \nonumber\\
&= k_BT\ln\left( \frac{(2N-k)!}{(2N-m-1)!}\frac{(k-1)!}{m!}\right) \nonumber\\
&\leq k_BT \bigg( (2N-k)\ln(2N-k)+(k-1)\ln(k-1)-(2N-m-1)\ln(2N-m-1)-m\ln(m) \nonumber \\
&\quad + \frac{1}{2} \ln\left(\frac{(2N-k)(k-1)}{(2N-m-1)m}\right) \bigg) + \mathcal{O}\left(\frac{1}{k}\right) \qquad \text{(Stirling approx.)} \nonumber\\
&=k_BT \bigg( (2N-k+1)\ln\left(\frac{2N-k}{k}\right) - (2N-m) \ln \left( \frac{2N-m}{m} \right) -2N \ln \left(\frac{m}{k}\right)  \nonumber\\
&\quad \underbrace{-(k-1)\ln \left( \frac{k}{k\!-\!1} \right)}_{=-1+\mathcal{O}\left(\frac{1}{k}\right)}+\underbrace{(2N\!-\!m\!-\!1) \ln \left(\frac{2N-m}{2N\!-\!m\!-\!1} \right)}_{=1+\mathcal{O}\left(\frac{1}{N}\right)}+\underbrace{\frac{1}{2}\ln\left ( \frac{2N\!-\!m\!-\!1}{2N-k} \frac{k-1}{m}\right )}_{\leq 0} \bigg) + \mathcal{O}\left(\frac{1}{k}\right) \nonumber\\
&\leq (2N-k+1)E_k - (2N-m) E_m -2N k_BT\ln \left(\frac{m}{k}\right) +\mathcal{O}\left(\frac{1}{k}\right)\label{sum1}.
\end{align}
Using (\ref{diff1}) and (\ref{finalsum}) we can already see that the last term in the first line of (\ref{var1}) goes as $\mathcal{O}\left( \frac{\ln^2 N}{N} \right)$ and thus disappears for $N \to \infty$. Now we rewrite the other terms,
\begin{align}
\begin{split}
\mathrm{Var}( W_N)&=(1-\alpha) \sum_{m=1}^{N-1} \bigg[E_m^2 q_m(1-q_m)+E_{m+1}^2 \norm{\pvec{m}{1}}(1-\norm{\pvec{m}{1}}) \\
&\label{variance} \quad +2E_{m+1}\left(\norm{\pvec{m}{1}} \ave{m} - \pvec{m}{1}\evec{m}{1} \right) \bigg]+ \mathcal{O}\left( \frac{\ln^2 N}{N} \right),\end{split}
\end{align}
where the change of indices does not matter since $E_N=p_0=0$. Writing the last two terms in the brackets explicitly, we get
\begin{align}
\norm{\pvec{m}{1}} \ave{m} \!-\! \pvec{m}{1}\evec{m}{1}\! &\overset{\eqref{pmwm}}{=}(1-\alpha) \sum_{k=1}^m \alpha^{m-k}q_k \left(\ave{m}-\ave{k-1}-(1-\norm{\pvec{k-1}{1}})E_k\right) \nonumber \\
&\hspace{-3mm}\overset{\eqref{pvec},\eqref{averagework}}{=}\!(1\!-\!\alpha) \sum_{k=1}^m \alpha^{m-k}q_k \left(\!\frac{1}{2N}\sum_{i=k}^m E_i \underbrace{(1\!-\!\alpha^{i})}_{\leq 1} \!-\! \left(\!1\!-\!\frac{k\!-\!1}{2N}\!+\!\frac{\alpha}{1\!-\!\alpha}\frac{1\!-\!\alpha^{k-1}}{2N}\right)E_k \right) \nonumber \nonumber\\
&\overset{\eqref{sum1}}{\leq}-(1-\alpha) \sum_{k=1}^m \alpha^{m-k}q_k \bigg((1-q_m)E_m+k_BT\ln\left(\frac{m}{k}\right) +\frac{\alpha}{1-\alpha}\frac{1}{2N}E_k\nonumber \\
&\quad + \ord{\frac{1}{kN}} + \ord{\alpha^k \frac{\ln N}{N} }\bigg)
\nonumber \\
&\overset{\eqref{pvec}}{=}\!-\!\norm{\pvec{m}{1}}(1-q_m)E_m\!-\!(1\!-\!\alpha)\sum_{k=1}^m\alpha^{m-k} q_k \left(k_BT\ln\left(\frac{m}{k}\right) +\frac{\alpha}{1-\alpha}\frac{1}{2N}E_k \right) \nonumber \\
&\quad +\ord{\frac{1}{N^2}}+\ord{m^2\alpha^m\frac{\ln N}{N^2}}.
\label{lastterms}
\end{align}
Next, we separately calculate the sum, using $-\ln(1+x)\leq -\left(x-\frac{x^2}{2}\right)$,
\begin{align}
&-(1-\alpha)k_BT\sum_{k=1}^m\alpha^{m-k} q_k \left(\ln\left(\frac{m}{k}\right) +\frac{\alpha}{1-\alpha}\frac{1}{2N}\ln\left(\frac{2N-k}{k}\right) \right) \nonumber \\
=&-(1-\alpha)k_BT\sum_{d=0}^{m-1} \alpha^{d}\  \frac{m-d}{2N} \left(\ln\left( \frac{m}{m-d} \right)+ \frac{\alpha}{1-\alpha} \frac{1}{2N} \ln \left(\frac{2N-m+d}{m-d}\right) \right) \nonumber \\
=&\!-\!(1\!-\!\alpha)k_BT\sum_{d=0}^{m-1}\! \alpha^{d} \frac{m-d}{2N} \!\left(\!\ln\!\left(\! 1\!+\!\frac{d}{m\!-\!d}\! \right)\!+\! \frac{\alpha}{1\!-\!\alpha} \frac{1}{2N} \!\left(\! \ln \!\left(\frac{2N\!-\!m}{m}\right)\!+\!\ln\! \left(\!\frac{2N\!-\!m\!+\!d}{2N-m}\right)\!+\!\ln \left(\frac{m}{m\!-\!d}\right)\!\right)\! \right) \nonumber \\
=&\!-\!(1\!-\!\alpha)k_BT\!\sum_{d=0}^{m-1}\! \alpha^{d} \frac{m\!-\!d}{2N} \!\left(\!\ln\!\left(\! 1\!+\!\frac{d}{m\!-\!d}\! \right)\!+\! \frac{\alpha}{1\!-\!\alpha} \frac{1}{2N} \!\left(\! \ln \!\left(\!\frac{2N\!-\!m}{m}\right)\!+\!\ln\! \left(\!1\!+\!\frac{d}{2N\!-\!m}\right)\!+\!\ln\!\left(\!1\!+\!\frac{d}{m\!-\!d}\right)\!\right)\! \right) \nonumber \\
\leq&\!-\!(1\!-\!\alpha\!)k_BT\!\sum_{d=0}^{m-1} \!\alpha^{d} \frac{m\!-\!d}{2N} \!\left(\!\frac{d}{m\!-\!d}\!-\!\frac{d^2}{2\!(m\!-\!d)^2} \!+\! \frac{\alpha}{1\!-\!\alpha} \frac{1}{2N} \!\left(\!\ln\!\left(\!\frac{2N\!-\!m}{m}\!\right)
\!+\!\ord{\frac{d}{N}} \!+\! \frac{d}{m\!-\!d} \!-\!\frac{d^2}{2(m\!-\!d)^2}\!\right)\! \right) \nonumber \\
\leq&-\frac{\alpha}{(1-\alpha)}\frac{k_BT}{2N}+\ord{\frac{m\alpha^m}{N}}+\underbrace{(1-\alpha)k_BT\sum_{d=0}^{m/2}\alpha^d \frac{d^2}{2Nm}}_{=\ord{\frac{1}{mN}}+\ord{\frac{m\alpha^{m/2}}{N}}}+\underbrace{(1-\alpha)k_BT\sum_{d=m/2}^{m-1}\alpha^{m/2}\frac{d^2}{4N}}_{=\ord{\frac{m^3\alpha^{m/2}}{N}}}\nonumber\\
& -k_BT\ln \left(\frac{2N-m}{m}\right) \frac{1}{4N^2}\frac{m \alpha}{(1-\alpha)} 
 +\ord{\frac{\ln N}{N^2}} +\ord{\frac{m}{N^3}} +\ord{\frac{1}{N^2}} +\ord{\frac{m^2a^m}{N^2}}  \nonumber \\
=&-\frac{\alpha}{(1-\alpha)}\frac{k_BT}{2N}-E_m\frac{m}{4N^2}\frac{\alpha}{(1-\alpha)}+\ord{\frac{1}{mN}}+\ord{\frac{m^3\alpha^{m/2}}{N}}+\ord{\frac{\ln N}{N^2}},
\end{align}
which we can insert into (\ref{lastterms}),
\begin{align}
\begin{split}
\norm{\pvec{m}{1}} \ave{m} - \pvec{m}{1}\evec{m}{1} &\leq-\norm{\pvec{m}{1}}(1-q_m)E_m-\frac{\alpha}{(1-\alpha)}\frac{k_BT}{2N}-E_m \frac{m}{4N^2}\frac{\alpha}{(1-\alpha)} \\
&\quad +\ord{\frac{1}{mN}}+\ord{\frac{m^3\alpha^{m/2}}{N}}+\ord{\frac{\ln N}{N^2}}.
\end{split}
\end{align}
Inserting this into the expression for the variance \eqref{variance} and rewriting the first line, we get
\begin{align}
\mathrm{Var}(W_N)&=(1\!-\!\alpha) \sum_{m=1}^{N-1} \bigg[(E_m^2\!+\!E_{m+1}^2) \norm{\pvec{m}{1}}(1\!-\!q_m) \!+\! \left(E_m^2(1-q_m)+E_{m+1}^2\norm{\pvec{m}{1}}\right)(q_m-\norm{\pvec{m}{1}})\nonumber\\
& \quad -2E_{m+1}\left( \norm{\pvec{m}{1}}(1-q_m)E_m+\frac{\alpha}{(1-\alpha)}\frac{k_BT}{2N}+E_m \frac{m}{4N^2}\frac{\alpha}{(1-\alpha)} \right) \bigg]+ \mathcal{O}\left( \frac{\ln^2 N}{N} \right) \nonumber \\
&=(1\!-\!\alpha) \sum_{m=1}^{N-1} \bigg[ \underbrace{(E_m\!-\!E_{m+1})^2  \norm{\pvec{m}{1}}(1\!-\!q_m) }_{=\ord{\frac{1}{m}}^2\cdot \ord{\frac{m}{N}}=\ord{\frac{1}{mN}}}+E_m^2 \underbrace{(q_m-\norm{\pvec{m}{1}})}_{=\frac{\alpha}{1-\alpha}\frac{1-\alpha^m}{2N}} - \!\underbrace{E_m^2( q_m-\norm{\pvec{m}{1}} )^2}_{=\ord{\ln N}^2 \ord{\frac{1}{N}}^2=\ord{\frac{\ln^2N}{N^2}}} \nonumber\\
& \quad \!+\!  \underbrace{(E_{m+1}^2\!-\!E_m^2)  \norm{\pvec{m}{1}} ( q_m\!-\!\norm{\pvec{m}{1}} )}_{=\ord{\frac{\ln N}{m}} \ord{\frac{m}{N}} \ord{\frac{1}{N}} = \ord{\frac{\ln N}{N^2}}} -2\! E_{m+1}\! \left(\! \frac{\alpha}{(1\!-\!\alpha)}\frac{k_BT}{2N}\!+\!E_m \frac{m}{4N^2}\frac{\alpha}{(1\!-\!\alpha)}\! \right)\! \bigg]\!+\! \mathcal{O}\!\left(\! \frac{\ln^2 N}{N}\! \right) \nonumber\\
&=\sum_{m=1}^{N-1} \bigg[\ord{\frac{1}{mN}}+E_m^2 \frac{\alpha}{2N} +\ord{\frac{\alpha^m\ln^2 N }{N}} -2E_{m}\left( \frac{\alpha k_BT}{2N}+E_m \frac{m \alpha}{4N^2} \right) \bigg]+ \mathcal{O}\left( \frac{\ln^2 N}{N} \right)\nonumber \\
&=\frac{\alpha}{2N} \sum_{m=1}^{N-1} \bigg[E_m^2\left(1- \frac{m }{2N}\right)-k_BTE_{m} \bigg]+ \mathcal{O}\left( \frac{\ln^2 N}{N} \right).
\end{align}
In the last step we need to calculate these two sums,
\begin{align}
-\frac{\alpha k_BT}{N}\sum_{m=1}^{N-1}E_{m}&=-\frac{\alpha (k_BT)^2}{N} \int_1^N \ln \left( \frac{2N-m}{m} \right) dm + \ord{\frac{\ln N}{N}} \nonumber\\
&=\frac{\alpha (k_BT)^2}{N}\big[ (2N-m)\ln (2N-m) +m \ln m \big]_1^N + \ord{\frac{\ln N}{N}} \nonumber\\
&= -2 \alpha (k_BT)^2\ln 2 + \ord{\frac{\ln N}{N}}
\end{align}
and
\begin{align}
\frac{\alpha}{2N}\sum_{m=1}^{N-1}E_m^2\left(1-\frac{m}{N}\right) &=\frac{\alpha(k_BT)^2}{2N}\int_1^{N}\ln^2 \left( \frac{2N-m}{m} \right) \left(1-\frac{m}{N}\right) dm + \ord{\frac{\ln^2 N}{N}} \nonumber \\
&=\frac{\alpha(k_BT)^2}{2N}\left[ -4N\ln(2N\!-\!m)\!+\!2m\ln \left( \frac{2N\!-\!m}{m} \right) \!+\!(1\!-\!\frac{m}{2N}) m \ln^2\left( \frac{2N\!-\!m}{m} \right)\right]_1^{N} \nonumber \\
& \quad + \ord{\frac{\ln^2 N}{N}} \nonumber \\
&= 2\alpha (k_BT)^2 \ln 2 + \ord{\frac{\ln^2 N}{N}}.
\end{align}
Thus, we get for the variance,
\begin{align}
\mathrm{Var}(W_N)= \ord{\frac{\ln^2 N}{N}} \overset{N\to \infty}{\longrightarrow} 0,
\end{align}
where we can see that for infinitely many steps $N$ the fluctuations vanish. This is in accordance with the Jarzynski equality \mbox{$e^{\beta \Delta F}=\langle e^{-\beta W} \rangle $}  \cite{Jarzynski97}, as for $N \to \infty$ we are looking at a quasistatic process with $\Delta F = -\langle W \rangle =F(\tau_{\rm B}^{(N)} ,H_{\rm S})-F(\tau_{\rm B}^{(0)} ,H_{\rm S}) $ and having no fluctuations implies $e^{-\beta \langle W \rangle}=\langle e^{-\beta W} \rangle $.

\section{Generalization to Arbitrary Systems }
\label{appGen}
We now extend our previous considerations to generic qudit systems and more general choices for the Hamiltonians of the bath systems. We consider a continuous family of Hamiltonians $H(s)$, $s \in [0,1]$, as well as the corresponding continuous family of thermal states $\tau(s)= \frac{e^{-\beta H(s)}}{\Tr[e^{-\beta H(s)}]}$, such that for a fixed $N$ we can draw the Hamiltonians $H^{(k)}_B=H(k/N)$ and the corresponding thermal states $\tau^{(k)}_B=\tau(k/N)$. The functions needs to be continuously differentiable and for optimal processes satisfy $\|\tau(0) -\rho_{\rm S}^{(0)} \|_1 \leq \delta\ll 1$, where $\delta = 0$ is always possible if the initial state $\rho_{\rm S}^{(0)}$ is a full-rank state.
We also define the continuous function,  
\begin{align}
F(\lambda):=F(\tau(\lambda), H_{\rm S}),
\end{align}
 with $\lambda \in (0,1)$. We take as a starting point the model of noise \eqref{modelnoise}. The derivation up to the second line of \eqref{generalwork} still follows naturally. 
 For simplicity, we will in the following first assume $\rho^{(0)}_{\rm S}=\tau_{\rm B}^{(0)}$ and hence $\delta=0$. In the case $\delta\neq 0$ we obtain an additional term, which we discuss below. To move on, we notice the identity, 
\begin{align}
\Tr\left((H_{\rm S} - H_{\rm B})(\tau - \rho) \right)&=F(\tau,H_{\rm S})-F(\rho,H_{\rm S})-F(\tau,H_{\rm B})+F(\rho,H_{\rm B}) \nonumber \\
&= F(\tau,H_{\rm S})-F(\rho,H_{\rm S})+TS(\rho \parallel \tau).
\end{align}
Using \eqref{recrel} in the case $\rho_{\rm S}^{(0)}=\tau_{\rm B}^{(0)}$, we can rewrite the second line of  \eqref{generalwork} as, 
\begin{align}\label{eq:deltazero}
W &= - (1-\alpha) \sum_{k=0}^{N-1} \Tr \left[ \left( H_{\rm S} - H_{\rm B}^{(k+1)} \right)  \left( \tau_{\rm B}^{(k+1)} - \sum_{j=0}^{k-1} (1-\alpha) \alpha^j \tau_{\rm B}^{(k-j)}-\alpha^k\tau_{\rm B}^{(0)} \right)  \right] \\
&= - (1-\alpha) \sum_{k=0}^{N-1} \Tr \left[ \left( H_{\rm S} \!-\! H_{\rm B}^{(k+1)} \right)  \left( \left(1\!-\!\alpha\right)\left( \frac{1\!-\!\alpha^{k} }{1\!-\!\alpha} \tau_{\rm B}^{(k+1)} \!-\! \sum_{j=0}^{k-1}  \alpha^j \tau_{\rm B}^{(k-j)}\right)\!+\!\alpha^k(\tau_{\rm B}^{(k+1)}\!-\!\tau_{\rm B}^{(0)} )\right)  \right]
\nonumber\\
&=- (1-\alpha)^2  \sum_{k=0}^{N-1}  \sum_{j=0}^{k-1} \alpha^j \Tr \left[ \left( H_{\rm S} - H_{\rm B}^{(k+1)} \right)  \left( \tau_{\rm B}^{(k+1)} -  \tau_{\rm B}^{(k-j)} \right)  \right] \nonumber\\
&\hspace{3mm} - (1-\alpha)\sum_{k=0}^{N-1} \alpha^{k} \Tr \left[  \left( H_{\rm S} - H_{\rm B}^{(k+1)} \right) \left(\tau_{\rm B}^{(k+1)} -\tau_{\rm B}^{(0)}\right)  \right] \nonumber\\
\begin{split}
&=-(1-\alpha)^2\sum_{k=0}^{N-1} \sum_{j=0}^{k-1} \alpha^j \left(F(\tau_{\rm B}^{(k+1)},H_{\rm S})-F(\tau_{\rm B}^{(k-j)},H_{\rm S})+TS(\tau_{\rm B}^{(k-j)} \parallel \tau_{\rm B}^{(k+1)}) \right)
\\
&\quad -(1-\alpha)\sum_{k=0}^{N-1}\alpha^k \left(F(\tau_{\rm B}^{(k+1)},H_{\rm S})-F(\tau_{\rm B}^{(0)},H_{\rm S})+ T S(\tau_{\rm B}^{(0)} \parallel \tau_{\rm B}^{(k+1)}) \right).
\end{split}
\label{expAppW}
\end{align}
In the case $\delta \neq 0$,  would have an additional term proportional $\alpha^k (\rho_{\rm B}^{(0)}-\tau_{\rm B}^{(0)})$ in \eqref{eq:deltazero} obtained by replacing $\rho_{\rm S}^{(0)}$ by $\tau_{\rm B}^{(0)} + \rho_{\rm S}^{(0)}-\tau_{\rm B}^{(0)}$. This leads to additional work
 \begin{align}
 W_0 &= (1-\alpha) \sum_{k=0}^{N-1} \alpha^k \Tr \left[ \left( H_{\rm B}^{(k)} - H_{\rm S} \right) \left( \tau_{\rm B}^{(0)} -\rho_{\rm B}^{(0)}  \right) \right] \nonumber \\
 &\leq  \delta   \max_s||H(s) - H_S||_\infty = \ord{\delta \ln \frac{1}{\delta}},
 \label{deltaneq0}
 \end{align}
where the last estimate follows from the fact that $\| H(0)\|_\infty=\ord{\ln(1/\delta)}$ is required to fulfill $\|\tau(0)-\rho_{\rm S}^{(0)}\|_1=\delta\ll 1$ and we assumed that $\max_s \| H(s) \|_\infty = \|H(0) \|_\infty$.
This fact will similarly lead to a divergence of the quantity $\int f(s) \mathrm{d} s$ below of the order $\log(1/\delta)$. Since we are interestead to leading order corrections, by choosing $\delta=1/N$ we can obtain a scaling as $\log(N)/N$ instead of $1/N$ in what follows. For simplicity, however, we concentrate on the case $\delta = 0$.

Since we are interested in leading order corrections up to order $1/N$, we have that
\begin{align}
F(\tau_{\rm B}^{(k)},H_{\rm S})-F(\tau_{\rm B}^{(k-j)},H_{\rm S})=\frac{j}{N}\dot{F}(k/N)+\frac{j^2}{2 N^2}\ddot{F}(k/N)+\mathcal{O}\left(\frac{j^3}{N^3}\right),
\label{Fappp}
\end{align} 
which will be useful to simplify \eqref{expAppW}. Similarly, we can expand $S(\cdot \parallel \cdot)$ as
 \begin{align}
S\left(\tau(\lambda+x) \parallel \tau(\lambda)\right) = \frac{x^2}{2} \frac{d^2 S(\tau(\lambda+x) \parallel \tau(\lambda))}{dx^2} \bigg|_{x=0} + \mathcal{O}(x^3),
\end{align}
which follows because $S(\rho \parallel \tau)$ has a minimum at $\rho=\tau$ and $S(\rho \parallel \rho)=0$, and define 
\begin{align}
f(\lambda)=T \frac{d^2 S(\tau(\lambda+x) \parallel \tau(\lambda))}{dx^2} \bigg|_{x=0}.
\label{flambda}
\end{align}
Although here we will not be concerned with the form of $f(\lambda)$, it is worth pointing out that  $f(\lambda)$ can be written as
\begin{align}
f(\lambda)= -\Tr(\dot{\tau}(\lambda) \dot{H}(\lambda)).
\end{align}

Let us now consider,
\begin{align}
W^{*}\!&=\!-(1\!-\!\alpha)^2\!\sum_{k=0}^{N-1} \sum_{j=0}^{k-1} \alpha^j \left(F(\tau_{\rm B}^{(k+1)}\!,H_{\rm S})\!-\!F(\tau_{\rm B}^{(k-j)}\!,H_{\rm S}) \right)\!-\!(1\!-\!\alpha)\!\sum_{k=0}^{N-1}\alpha^k \left(F(\tau_{\rm B}^{(k+1)}\!,H_{\rm S})\!-\!F(\tau_{\rm B}^{(0)}\!,H_{\rm S})\! \right)
\nonumber\\
\mathcal{E}&=-W+W^{*}
\end{align}
We now use  \eqref{Fappp} to simplify  $W^{*}$ and work out the sum in $j$. Neglecting terms of order $\ord{\alpha^N}$, we obtain
\begin{align}
W^*&=\!-\!\sum_{k=0}^{N-1} \left[ (1\!-\!\alpha)^2 \sum_{j=0}^{k-1} \alpha^j  \left(\frac{j+1}{N}\dot{F}(k/N)\!+\!\frac{(j+1)^2}{2 N^2}\ddot{F}(k/N) \right)+(1\!-\!\alpha) \alpha^k \frac{k+1}{N} \dot{F}(k/N)\right]\!+\!\mathcal{O}(1/N^2) \nonumber \\
&=-\sum_{k=0}^{N-1} \left[ (1-\alpha^{k+1}) \frac{\dot{F}(k/N)}{N} +(1-\alpha)^2\sum_{j=0}^{k-1} \alpha^j\frac{(j+1)^2}{2 N^2}\ddot{F}(k/N)\right]+\mathcal{O}(1/N^2)\nonumber\\
&=\!-\!\sum_{k=0}^{N-1}\! \left[ \!\left(\! \frac{\dot{F}(k/N)}{N}\!+\!\frac{\ddot{F}(k/N)}{2N^2}\right)\!-\!\alpha^{k+1} \frac{\dot{F}(k/N)}{N} \!+\!\frac{\ddot{F}(k/N)}{2N^2}\!\left(\!-\!1\!+\!(1\!-\!\alpha)^2\sum_{j=0}^{k-1} \alpha^j(j\!+\!1)^2\!\right)\!\right]\!+\!\mathcal{O}(1/N^2)
\nonumber\\
&=\Delta F -\sum_{k=0}^{N-1} \left[-\alpha^{k+1} \frac{\dot{F}(k/N)}{N} +\frac{\ddot{F}(k/N)}{2N^2}\left(-1+(1-\alpha)^2\sum_{j=0}^{k-1} \alpha^j(j+1)^2\right)\right]+\mathcal{O}(1/N^2).
\end{align}
Using that $\dot{F}(k/N)=\dot{F}(0_+)+\mathcal{O}(k/N)$, and again neglecting elements with $\ord{\alpha^N}$, we simplify the expression to
\begin{align}
W^*\!&=\Delta F + \dot{F}(0_+)\frac{\alpha}{N(1\!-\!\alpha)}\!-\!\sum_{k=0}^{N-1} \frac{\ddot{F}(k/N)}{2N^2} \frac{2 \alpha \!+\!\left(2 k^2\!+\!2 k\!-\!1\right) \alpha ^{k+1}\!-\!k^2 \alpha ^{k+2}\!-\!(k+1)^2 \alpha
   ^k}{1-\alpha }+\mathcal{O}(1/N^2)
   \nonumber\\
&=\Delta F-\frac{\alpha}{N(1-\alpha)}\left(- \dot{F}(0_+)+\frac{1}{N}\sum_{k=0}^{N-1}\ddot{F}(k/N) \right)+\mathcal{O}(1/N^2)
   \nonumber\\
&=\Delta F-\frac{\alpha}{N(1-\alpha)}\left(- \dot{F}(0_+)+\int_0^1 ds \ddot{F}(s) \right)+\mathcal{O}(1/N^2)
   \nonumber\\
&=\Delta F-\frac{\alpha}{N(1-\alpha)}\left(\dot{F}(1)- 2\dot{F}(0_+)\right)+\mathcal{O}(1/N^2) \label{Wstar}
\end{align}
On the other hand, we have for $\mathcal{E}$
\begin{align}
\mathcal{E}&=(1-\alpha)^2\sum_{k=0}^{N-1} \sum_{j=0}^{k-1} \alpha^j TS(\tau_{\rm B}^{(k-j)} \parallel \tau_{\rm B}^{(k+1)})  +(1-\alpha)\sum_{k=0}^{N-1}\alpha^k T S(\tau_{\rm B}^{(0)} \parallel \tau_{\rm B}^{(k+1)}) 
\nonumber\\
&=\frac{1}{2N^2}\left(\sum_{k=0}^{N-1} f(k/N) (1-\alpha)^2\sum_{j=0}^{k-1} \alpha^j  (j+1)^2 \right)+\mathcal{O}(1/N^2)
\nonumber\\
&=\frac{1}{2N^2}\left(1+\frac{2\alpha}{1-\alpha} \right)\sum_{k=0}^{N-1} f(k/N)+\mathcal{O}(1/N^2)
\nonumber\\
&=\frac{1}{2N}\left(1+\frac{2\alpha}{1-\alpha} \right)\int_0^1 ds f(s)+\mathcal{O}(1/N^2) \nonumber \\
&=: \frac{\Gamma}{N} \left(1 + \frac{2\alpha}{1-\alpha}\right)+\mathcal{O}(1/N^2) .\label{epsilon}
\end{align}

\section{Isothermal Processes and Thermal Maps}
\label{appframe}
In this section, we provide the proof of the results presented in \ref{resultsgibbsmaps}.
Before starting the formal proof, let us discuss the main ideas that go into the proof. 
For a fixed path of Hamiltonians, we are dealing with a process of a large number $N$ of steps. 
For a sufficiently large number of steps, we can expect that i) the unitary time-evolution between two thermalization steps converges to the time-evolution generated by a constant Hamiltonian and  ii) that the state of the system remains relatively close to a thermal state during a protocol, even though each individual thermalizing map is imperfect and the system therefore "lags behind" the actual thermal state.
To simplify interpretation, we therefore split the total dissipated work into three parts:
\begin{align}
	\Delta F_{\mathrm{iso}} - W_{\mathrm{iso}} = \gamma + \varepsilon + \kappa.
\end{align}
Here, $\gamma$ is the error that would arise in the case of  perfect thermalizations and we would replace the unitary time-evolution inbetween the thermalization steps by instantaneous quenches. It is imply due to the fact that we consider a finite protocol. 
The term $\varepsilon$ vanishes in the case of perfect thermalizations and the term $\kappa$ vanishes when we consider instantaneous quenches instead of unitary time-evolution.
We will show that 
\begin{align}
	\gamma &= \frac{\Gamma}{N}  + \mathcal{O}\left(\frac{1}{N^2}\right),\quad
	\varepsilon = \frac{\alpha}{1-\alpha}\left(\frac{\Lambda}{N} + \mathcal{O}\left(\frac{1}{N^{2}}\right)\right),\quad
	\kappa = \frac{K}{N} + \mathcal{O}\left(\frac{1}{N^2}\right).
\end{align}

In the following we will use the simplified notation 
\begin{align}
	U_{i} \coloneqq U_{t_{i},t_{i-1}}
\end{align}
for the unitary that is generated by $H(t)$ from time $t_{i-1}$ to $t_{i}$. 
We remind the reader that we assume
\begin{align}
	\delta t = t_{i}-t_{i-1} = 1/N.
\end{align}
Recall that the total work that can be extracted during the protocol
is given by
\begin{align}
	W_{\mathrm{iso}} =	\sum_{i=1}^N\Tr\left( H^{(i-1)} \sigma^{(i-1)} - H^{(i)}\rho^{(i)} \right)
	= \sum_{i=1}^N \Tr\left( H^{(i-1)} \sigma^{(i-1)} - H^{(i)} U_{i}\sigma^{(i-1)} U_{i}^\dagger \right).
\end{align}
Adding and substracting zeroes allows us to re-write the work as
\begin{align}
	W_{\mathrm{iso}} &= \sum_{i=1}^N\Tr\left(\Delta H^{(i)} \tau^{(i-1)}\right)\nonumber \\
	    &+\ \sum_{i=1}^N\Tr\left(\Delta H^{(i)} \left(\sigma^{(i-1)}-\tau^{(i-1)}\right)\right)\nonumber \\
	    &+\ \sum_{i=1}^N\Tr\left(H^{(i)}\left(\sigma^{(i-1)} - U_i \sigma^{(i-1)}U_i^\dagger \right)\right),
\end{align}
where $\Delta H^{(i)} = H^{(i-1)} - H^{(i)}$. Each of the terms corresponds to one of the terms for the dissipated work above. In the following we will bound each term separately. We will assume that both the norm of the Hamiltonian as well as the norm of the derivative of the Hamiltonian along the Hamiltonian trajectory are bounded.

\subsection{First Term}
The first term has previously been shown to yield (see, e.g., \cite{Anders13,AbergWork,Gallego14,Jacobs2009})
\begin{align}
	\sum_{i=1}^N\Tr\left(\Delta H^{(i)} \tau^{(i-1)}\right) = \Delta F_{\mathrm{iso}} - \frac{\Gamma}{N} - \mathcal{O}\left(\frac{1}{N^2}\right),
\end{align}
where $\Gamma$ is given by \cite{scandi2018thermodynamic} (also see Appendix \ref{appGen})
\begin{align}
	\Gamma = -\frac{1}{2} \int_0^1 ds \Tr\left(\dot \tau(s) \dot H(s)\right).
\end{align}
\subsection{Second Term}
The second term vanishes in the case of perfect thermalization, it hence corresponds to the dissipation $\varepsilon$. To obtain the bound,we first emphasize that $\Delta H^{(i)}$ is of order $1/N$. 
We therefore essentially have to show that $\sigma^{(i-1)}-\tau^{(i-1)}$ is also of order $1/N$. 
For notational reasons, it is useful to introduce the quantities
\begin{align}
	\left\| \partial H^{(i)} \right\|_\infty &\coloneqq \max_{t\in[t_{i-1},t_i]} \| H(t) - H^{(i)}\|_\infty = \mathcal{O}(1/N),\\
	 \|  \partial H \|_\infty &\coloneqq \max_i \left\| \partial H^{(i)} \right\|_\infty = \mathcal{O}(1/N). 
\end{align}
Note that here we use the Schatten norms, i.e. $\left\| A \right \|_p = \left( \sum_{n\geq 1} s_n^p(A)\right)^{1/p}$, where $s_n^p(A)$ denotes the $n$-th singular values of $A$. 
In particular this implies that $\left\| A \right \|_\infty$ corresponds to the operator norm and therefore assigns the largest singular value of $A$, while $\left\| A \right \|_1 = \Tr \left| A \right|$ corresponds to the trace norm.
We can then bound the second term using H\"older's inequality as
\begin{align}
	\sum_{i=1}^N\Tr\left(\Delta H^{(i)} \left(\sigma^{(i-1)}-\tau^{(i-1)}\right)\right) &\geq -\sum_{i=1}^N \|\partial H\|_\infty \| \sigma^{(i-1)}-\tau^{(i-1)} \|_1.
\end{align}
We can now make use of the fact that the thermalizing maps have a thermalization parameter $\alpha$ to write
\begin{align}
	\norm{\sigma^{(i)} - \tau^{(i)}} &\leq \alpha \norm{U_i \sigma^{(i-1)} U_{i}^\dagger - \tau^{(i)}} \nonumber \\
	&\leq \alpha \norm{\sigma^{(i-1)} - U_{i}^\dagger \tau^{(i)}  U_i} \leq \alpha \left(\norm{\sigma^{(i-1)} - \tau^{(i)}} + \norm{U_i^\dagger \tau^{(i)} U_i - \tau^{(i)}}\right).\label{eq:stateapprox} 
\end{align}
At this point we have to make use of the fact that the unitary $U_i$ is close to that generated by the Hamiltonian $H^{(i)}$. To increase clarity of presentation, we delegate the proof of this fact to Section~\ref{app:unitaries}. Using the notation 
\begin{align}
	\tilde U_i \coloneqq \e^{-\mathrm{i} H^{(i)}\delta t},\quad \delta U_i \coloneqq U_i - \tilde U_i,
\end{align}
we show in Section~\ref{app:unitaries} that
\begin{align}
	\label{eq:deltaUbound}
	\| \delta U_i \|_\infty = \mathcal{O}\left(1/N^2\right).
\end{align}

Since $[\tau^{(i)}, \tilde U_i]=0$, we can use this result to estimate the last term in \eqref{eq:stateapprox} as
\begin{align}
	\norm{U_i^\dagger \tau^{(i)} U_i - \tau^{(i)}} = \norm{U_i^\dagger \tau^{(i)} U_i - \tilde U_i^\dagger \tau^{(i)}\tilde U_i} \leq 2 
	\| \delta U_i\|_\infty  = \mathcal O(1/N^2), 
\end{align}
where we made use of H\"older's inequality.
Adding and subtracting zero and using the triangle inequality again, we then arrive at
\begin{align}
	\norm{\sigma^{(i)} - \tau^{(i)}} &\leq \alpha \left(\norm{\sigma^{(i-1)} - \tau^{(i-1)}} + \norm{\tau^{(i-1)}-\tau^{(i)}} + \mathcal O(1/N^2) 
	\right) 
\end{align}
We now bound the second term in this expression. To do this, we first use Pinsker's inequality $\| \rho-\sigma\|_1 \leq \sqrt{2 S(\rho \| \sigma)}$, where $S(\cdot \| \cdot)$ is again the quantum relative entropy.
Using the results around \eqref{flambda}, we then find
\begin{align}
	\norm{\tau^{(i)} - \tau^{(i-1)}}&\leq \sqrt{2 S(\tau^{(i)} \| \tau^{(i-1)})} = \sqrt{ \delta t^2 f\left(\frac{i}{N}\right)+ \mathcal{O}(\delta t^3)} \nonumber \\
	&=\delta t \sqrt{f\left(\frac{i}{N}\right)} + \mathcal{O}\left(\delta t^2\right)
	\leq \frac{1}{N} \sqrt{f_{\rm max}} +\mathcal O\left(\frac{1}{N^{2}}\right) ,
\end{align}
where we defined $f(\lambda)$ as in \eqref{flambda} and $f_{\rm max} = \max_{\lambda} f(\lambda)$. 
This yields
\begin{align}
	\norm{\sigma^{(i)} - \tau^{(i)}} &\leq \alpha \left(\norm{\sigma^{(i-1)} - \tau^{(i-1)}} +\frac{\sqrt{f_{\rm max}}}{N} + \mathcal O(1/N^{2}) 
	\right) \nonumber \\
	&\leq \sum_{k=1}^{i}\alpha^k \left(\frac{\sqrt{f_{\rm max}}}{N} + \mathcal O(1/N^{2})\right) + \alpha^{i}\| \sigma^{(0)} - \tau^{(0)} \|_1 \nonumber \\
	&= \frac{\alpha(1-\alpha^i)}{1-\alpha}\left(\frac{\sqrt{f_{\rm max}}}{N} + \mathcal O(1/N^{2})\right)+ \alpha^i \| \sigma^{(0)} - \tau^{(0)}\|_1.
\end{align}
Combining all the above bounds, we finally obtain
\begin{align}
	\sum_{i=1}^N\Tr\left(\Delta H^{(i)} \left(\sigma^{(i-1)}-\tau^{(i-1)}\right)\right) &\geq -\sum_{i=1}^N \|\partial H\|_\infty \| \sigma^{(i-1)}-\tau^{(i-1)} \|_1 \nonumber \\
	&\geq \frac{\alpha}{1-\alpha}\| \partial H\|_\infty \left(\sqrt{f_{\rm max}} +  \mathcal O(1/N^{1/2})+ \| \sigma^{(0)}- \tau^{(0)}\|_1 \right) \nonumber \\
	&\geq \frac{\alpha}{1-\alpha}\left(\frac{\Lambda}{N}+ \mathcal O\left(1/N^{2}\right)\right),
\end{align}
for a constant $\Lambda$.

\subsection{Third Term}
The third term vanishes in the case where unitary time-evolution is replaced by instantaneous quenches, so that $U_i = \mathbbm{1}$. It hence corresponds to the dissipation $\kappa$.
To bound it, we again make use of the approximation of the unitaries $U_i$ by the time-evolution under $H^{(i)}$, given by $\tilde U_i$. 
Since $[\tilde U_i, H^{(i)}]=0$, we can make use of $U_i = \tilde U_i + \delta U_i$ and bound this term using H\"older's inequality as
\begin{align}
	\sum_{i=1}^N\Tr\left(\left(H^{(i)} - U_i^\dagger H^{(i)} U_i\right)\sigma^{(i-1)}\right)&\geq -\sum_{i=1}^N \| H^{(i)} - U_i^\dagger H^{(i)} U_i \|_\infty \nonumber \\
	&\geq - \| H\|_\infty \sum_{i=1}^N 2\| \delta U_i\|_\infty 
	\geq -\frac{K}{N} + \mathcal{O}(1/N^2),
\end{align}
where $K$ is a constant determined by the trajectory of Hamiltonians and where we introduced 
\begin{align}
	\| H \|_\infty &\coloneqq \max_t \|H(t)\|_\infty.
\end{align}
The second inequality follows from
\begin{align}
	\| H^{(i)} - U_i^\dagger H^{(i)} U_i \|_\infty &= \| \tilde U_i^\dagger H^{(i)} (U_i - \delta U_i) - (\tilde U_i^\dagger+\delta U_i^\dagger) H^{(i)} U_i) \|_\infty \nonumber \\
	&\leq \|\tilde U_i^\dagger H^{(i)}\delta U_i\|_\infty  + \| \delta U_i^\dagger H^{(i)} U_i\|_\infty \nonumber \\
	&\leq 2 \| H^{(i)}\|_\infty \| \delta U_i\|_\infty,
\end{align}
where in the last step we used $\|A B\|_\infty \leq \|A\|_\infty \| B\|_\infty$, $\|A^\dagger \|_\infty = \|A \|_\infty$ and $\| U\|_\infty=1$ for any unitary $U$.

\subsection{Approximation of unitaries}
\label{app:unitaries}
In this section we show that the unitaries $U_i$ only differ from $\tilde U_i := \e^{-\mathrm{i} H^{(i)} \delta t}$ by an amount of order $\mathcal{O}(\delta t^2)$. 

To do that, we first write out the time-ordered exponential that defines $U_i$:
\begin{align}
U_i=\mathcal{T}\e^{-i\int_{t_{i-1}}^{t_i} H(t) dt}
=\sum_{k=0}^\infty \frac{(-i)^k}{k!} \int_{t_{i-1}}^{t_i}\mathrm{d}t_{i_1}\cdots \int_{t_{i-1}}^{t_{i}} \mathrm{d}t_{i_k} \mathcal{T} \{ H(t_{i_1}) \cdots  H(t_{i_k})  \}.
\end{align}
Bringing $\tilde{U}_i$ into a similar form we get
\begin{align}
\tilde{U}_i=\sum_{k=0}^\infty \frac{(-i)^k}{k!} (H^{(i)}\delta t)^k=\sum_{k=0}^\infty \frac{(-i)^k}{k!} {H^{(i)}}^k (t_i-t_{i-1})^k= \sum_{k=0}^\infty \frac{(-i)^k}{k!} {H^{(i)}}^k \int_{t_{i-1}}^{t_i}  \mathrm{d}t_{i_1} \cdots \int_{t_{i-1}}^{t_i}  \mathrm{d}t_{i_k}.
\end{align}
We can then upper bound their difference as
\begin{align}
\|\delta U_i\|_\infty =	\| U_i -\tilde{U}_i \|_\infty &\leq \sum_{k=1}^\infty \frac{1}{k!} \int_{t_{i-1}}^{t_i}  \mathrm{d}t_{i_1} \cdots \int_{t_{i-1}}^{t_i}  \mathrm{d}t_{i_k}  \left\| \mathcal{T} \left\{ H(t_{i_1}) \cdots H(t_{i_k}) \right\} - {H^{(i)}}^k \right\|_\infty \nonumber \\
&\leq \sum_{k=1}^\infty \frac{1}{k!} \delta t^k \left(k \| H \|_\infty^{k-1} \| \partial H^{(i)} \|_\infty\right) \nonumber \\
&\leq \e^{\| H\|_\infty \delta t} \delta t\, \| \partial H\|_\infty = \mathcal{O}(1/N^2).
\end{align}
Here, the second inequality follows from a telescope-like sum by iterating the bound
\begin{align}
	&\left\| H(t_{i_1})\cdots H(t_{i_k}) - {H^{(i)}}^k\right\|_\infty \\
	&\qquad \leq 
	\left\| H(t_{i_1})\cdots H(t_{i_k})- H(t_{i_1})\cdots H(t_{i_{k-1}})H^{(i)}\right\|_\infty + \left\| H(t_{i_1})\cdots H(t_{i_{k-1}})H^{(i)} - {H^{(i)}}^{k}\right\|_\infty\nonumber\\ 
	&\qquad\leq \left\| H(t_{i_1}) \right\|_\infty \cdots \left\| H(t_{i_{k-1}})\right\|_\infty \left\| H(t_{i_k}) - H^{(i)} \right\|_\infty 
 	+ \left\| H^{(i)}\right\|_\infty \left\|H(t_{i_1})\cdots H(t_{i_{k-1}}) - {H^{(i)}}^{k-1}\right\|_\infty \nonumber\\
	&\qquad\leq \| H\|^{k-1}_\infty \left\|\partial H^{(i)}\right\|_\infty + \| H\|_\infty  \left\|H(t_{i_1})\cdots H(t_{i_{k-1}}) - {H^{(i)}}^{k-1}\right\|_\infty.
\end{align}
This finishes the proof.

\end{document}